\documentclass[12pt,a4paper]{article}
\usepackage{graphicx}
\usepackage{amssymb}
\usepackage{amsmath}
\usepackage{bm}
\usepackage{color}
\usepackage{theorem}
\usepackage{subfigure}
\usepackage{amsfonts}

\usepackage[sort&compress,numbers, merge]{natbib}

\definecolor{Orange}{cmyk}{0,0.61,0.87,0}
\definecolor{JungleGreen}{cmyk}{0.99,0,0.52,0}
\definecolor{OliveGreen}{cmyk}{0.64,0,0.95,0.40}
\definecolor{Brown}{cmyk}{0,0.81,1,0.60}
\definecolor{RoyalBlue}{cmyk}{0.71,0.53,0,0.12}

\newcommand{\be}{\begin{equation}}
\newcommand{\ee}{\end{equation}}
\newcommand{\bea}{\begin{eqnarray}}
\newcommand{\eea}{\end{eqnarray}}
\newcommand{\eq}[1]{Eq.~(\ref{#1})}
\newcommand{\la}{\langle}
\newcommand{\ra}{\rangle}

\usepackage[colorlinks=True, citecolor=blue, linkcolor=blue, urlcolor=black]{hyperref}

\allowdisplaybreaks[1]

\newcommand{\Slash}[1]{{\ooalign{\hfil/\hfil\crcr$#1$}}}

\begin{document}

\begin{titlepage}
\begin{center}
\hfill TUM-HEP-1260/20

\vspace{2.0cm}
{\Large\bf  
Renormalization 
\vspace{.3cm}
of Higher-Dimensional Operators\\
from On-shell Amplitudes
}

\vspace{1.0cm}
{\small \bf 
Pietro Baratella$^{a}$,
Clara Fernandez$^{b}$ and
Alex Pomarol$^{b,c}$
}

\vspace{0.7cm}
{\it\footnotesize
${}^a$Technische Universit\"{a}t M\"{u}nchen, Physik-Department, 85748 Garching\\
${}^b$IFAE and BIST, Universitat Aut\`onoma de Barcelona, 08193 Bellaterra, Barcelona\\
${}^c$Departament de F\'isica, Universitat Aut\`onoma de Barcelona, 08193 Bellaterra, Barcelona\\
}

\vspace{0.9cm}
\abstract

On-shell amplitude methods allow to 
 derive   one-loop renormalization  effects
 from just tree-level amplitudes, with no need of  loop calculations.
We derive  a simple formula to obtain the  anomalous dimensions  of higher-dimensional operators
from a product of tree-level amplitudes.
 We show how this  works for dimension-6 operators of the  Standard Model,
providing explicit examples of  the simplicity, elegance and  efficiency of the method.
Many anomalous dimensions  can be calculated from the same Standard Model tree-level amplitude, 
displaying the attractive recycling aspect of the on-shell method. 
With this method, it is possible 
 to relate anomalous dimensions  that in  the Feynman approach
arise from  very  different diagrams, and   obtain   non-trivial checks of their relative coefficients.
We compare our results to those
in   the  literature, where  ordinary methods have been applied.

\end{center}
\end{titlepage}
\setcounter{footnote}{0}


\section{Introduction}

Effective Field Theories (EFT)  are  useful tools  to  describe  the relevant physics emerging 
at some given low-energy scale.
EFTs are usually  defined via  Lagrangians, whose terms or local operators ${\cal O}_i$ are organized
according to  an expansion in  derivatives and fields over a mass scale $\Lambda$.
This scale $\Lambda$  is believed to be associated with some new physics scale, above which new degrees of freedom must be incorporated into the theory.
The virtue of  an EFT is that, for low-energy experiments, with $E\ll\Lambda$,  only a few operators  are relevant,  those with  the lowest possible dimension, with 
  higher-dimensional operators  bringing  only small corrections, 
  as they are suppressed by  powers of $E/\Lambda$.
  
  Although small, the  effects from 
higher-dimensional operators  are  of crucial interest.
In the Standard Model (SM), for instance,
 higher-dimensional operators   provide indirect imprints of new physics.
For this reason, a lot of effort has been devoted to understand their  impact in low-energy experiments.

At the quantum level, operators of equal dimension mix with each other. 
This mixing is encoded in the anomalous dimensions of the corresponding Wilson coefficients $C_{{\cal O}_i}$,
which are defined through $\Delta {\cal L}=\sum_i C_{{\cal O}_i} {\cal O}_i$. The anomalous dimensions $\gamma_{i}$ are given by
\be
 \gamma_i\equiv \frac{dC_{{\cal O}_i}}{d\ln\mu}=\sum_j\gamma_{ij}\, C_{{\cal O}_j}\,,
\label{defAD}
\ee
where $\mu$ is the renormalization scale. The calculation of $\gamma_{i}$  in the SM EFT
is important to understand how experiments can determine or constrain the different 
Wilson coefficients, especially when  the energy scale of the experiment is much smaller than $\Lambda$.

We would like to follow here an alternative approach 
based on on-shell amplitude methods.
In this approach, a theory is defined by its particle content and  certain ``building-block" on-shell amplitudes,
with no need of Lagrangians.
As in the standard EFT procedure, we can also organize these
building-block amplitudes  in an expansion in $E/\Lambda$, and study their mixing 
via quantum loops. By requiring the amplitudes to be independent of the renormalization scale,
one can obtain  the analogue of the anomalous dimensions  $\gamma_{i}$
of  \eq{defAD}. In this case,  the role of the Wilson coefficients 
$C_{{\cal O}_i}$  is played by  the coefficients in front   of the buiding-block amplitudes, as we will describe below in detail.

One important advantage of working with on-shell amplitudes  is that this set-up naturally allows us to implement
  generalized unitarity methods,
extensively developed in the literature in recent years  \cite{Dixon:2013uaa},
to obtain $\gamma_{i}$ without the need of performing loop calculations.
Indeed, 
the divergencies of one-loop amplitudes can be obtained from
products of tree-level amplitudes (integrated over some phase space),
making the determination of the anomalous dimension quite simple.

We will mainly concentrate here in amplitudes at  order $E^2/\Lambda^2$ and consider
only massless states.
Moreover, we will restrict  to cases in which  IR divergencies are absent, 
and show how their cancellation 
allows to extract
anomalous dimensions from  double unitarity cuts of the one-loop amplitude,
with no need of any further cut.
This provides a simpler way to calculate anomalous dimensions than 
previously reported in Refs.~\cite{ArkaniHamed:2008gz,Caron-Huot:2016cwu}.

One of the main purposes of this article is to analyze the advantages or 
disadvantages of the  on-shell method versus the ordinary Feynman approach, especially 
in  cases  of phenomenological interest.
For this reason, we will  present  in detail the 
calculation of   the anomalous dimensions
of certain dimension-6 operators of the SM.
In particular, we will look at the dipole SU(2)$_L$ operator of the electron,
and calculate all contributions to its anomalous dimension.

We will see that the method is quite efficient,
as it essentially only  requires the calculation of a  few SM   amplitudes,
apart from some trivial angular integration.
Moreover, we will see that the same
SM amplitudes allow to calculate many other anomalous dimensions of the SM EFT.
This will show the  ``recycling" advantage   of on-shell methods,
where new calculations are obtained from previous ones, with no need to start the calculation from
the beginning, as it is usual in the Feynman diagrammatic approach.
This will also allow  to relate    $\gamma_i$
that originate from very different Feynman diagrams,
providing    non-trivial checks of previous results in the literature.

The article is organized in the following way.
In Section~\ref{sec1} we  present  what we call the building-block amplitudes of effective theories 
at order $E^2/\Lambda^2$.
In Section~\ref{sec2}
 we derive a formula to calculate  one-loop UV divergencies from tree-level amplitudes, 
 and relate it to previous ones obtained  in the literature.
 In Section~\ref{smcal} we use the formula to calculate 
   the anomalous dimensions of the dimension-6 dipole operator of the SM.
  We also show  the  correlation with the anomalous dimensions of  $\psi^4$ operators.
 In Section~\ref{sec:conclusion} we provide some conclusions.
We implement the article with  four Appendices.
In Appendix~\ref{app1}
we show how the cancellation of IR divergencies  leads to the 
 absence of triangle and box contributions in the sum over  the  double cuts of an amplitude, at least  at the order we are interested in.
In Appendix~\ref{app2} we  provide our conventions,  and 
derive some  SM amplitudes that are used  in the calculation of the anomalous dimensions.
In Appendix~\ref{app3} we 
relate our building-block amplitudes to  operators in the SM EFT Lagrangian
and provide a  dictionary between them.
Finally, in Appendix~\ref{app4} we briefly 
extend our analysis to dimension-5 operators.

\section{Effective Theories via on-shell amplitudes}
\label{sec1}

In the on-shell amplitude approach, a theory is  defined from its 
particle content and  scattering amplitudes.
All amplitudes can be constructed from lower-point ones,
and the lowest-point amplitudes play the role of building-blocks of the theory.

{As anticipated}, we will consider theories with only massless states, 
and classify the scattering amplitudes by their number of external legs $n$ 
and total helicity $h$, with all scattering states chosen to be incoming.
To write down amplitudes, we will use  the spinor-helicity notation
\cite{Dixon:2013uaa}, where  momenta and polarizations are written as  product of spinors
$|i\ra_\alpha$ and $|j]^{\dot\alpha}$, of helicity $h=-1/2$ and $h=1/2$ respectively. 
Our conventions are found  in Appendix \ref{app2}.
The purpose of spinor-helicity variables is to efficiently implement Poincar\'e covariance of scattering amplitudes. 
 The most important property, which is enforced by the little group, is
that amplitudes involving a state $i$
 of helicity $h$ must contain 
 the spinors $|i\ra$ and $|i]$ in such a way that the power of $|i]$ {minus}
 the power of $|i\ra$  equals  $2h$. 
Lorentz invariance imposes that spinors must appear in contractions $\la ij\ra$ or $[ij]$.
This makes the determination of amplitudes quite straightforward.

When the theory is also invariant under some internal symmetry group,
amplitudes behave as invariant tensors under its action on particle multiplets. 
In this section we will not bother to specify the form of group-tensors, reducing to the so-called
``color-stripped'' amplitudes \cite{Dixon:2013uaa}.
In Section~\ref{smcal} we will however consider explicit examples for SM amplitudes, and  
the invariant tensors will be provided. 
Several SM examples can also be found in Refs.~\cite{Christensen:2018zcq,Shadmi:2018xan,Ma:2019gtx}.

Similarly as it is done for operators, 
we can consider  the building-block amplitudes that define the theory 
as organized according to  an  expansion in  $E/\Lambda$,
which means an expansion in  powers of $\la ij\ra/\Lambda$ and $[ij]/\Lambda$.
When  we go beyond the ordinary interactions arising from dimensionless couplings (the equivalent of dimension-4 operators),  we find now extra interactions  at any order in $E/\Lambda$.
Since we will pay special attention to applications in the SM, we will concentrate here in $E^2/\Lambda^2$ terms, which are the leading corrections to the SM when
lepton number is conserved. 
We  leave for  Appendix~\ref{app4}  the discussion on terms of order $E/\Lambda$.

For a generic theory of (\textit{i}) vector bosons 
 $V_\pm$ with  helicity $h=\pm1$, (\textit{ii}) Weyl fermions $\psi$  with $h=-1/2$, and (\textit{iii}) scalars $\phi$,
 we have  the following building-block amplitudes  at order $E^2/\Lambda^2$ (up to complex conjugation):
\begin{itemize}
\item 
{\bf n=3:} 
\be
{\cal A}_{F^3}(1_{V_-},2_{V_-},3_{V_-})=\frac{C_{F^3}}{\Lambda^2}\la 12\ra \la 23\ra \la 31\ra \,,
\label{F3d}
\ee
that has  $h=-3$.  It is quite straightforward to see that this is the only amplitude at $n=3$.
Since  $n=3$ amplitudes have mass  dimension one, they 
must contain 3 powers of either brackets $\la ij\ra$ or squares $[ij]$ in the numerator. 
Moreover, we have the condition $\la ij\ra  [ji]=2p_i\cdot p_j=0$ ($i,j=1,2,3$), that forces the vanishing of either all $[ij]$,
in which case we can only have \eq{F3d}, or all $\la ij\ra$,  that leaves its complex-conjugated version as the only possibility.
It is important to notice that \eq{F3d}
is antisymmetric 
under $i\leftrightarrow j$, and can only arise for non-abelian gauge bosons,  
in which case the full amplitude is proportional to the structure constants.

\item 
{\bf n=4:} 
These amplitudes are dimensionless, so they must contain 2 powers of brackets or squares. 
We have the following possibilities, with total helicity $h=-2$:
\bea
{\cal A}_{F^2\phi^2}(1_{V_-},2_{V_-},3_{\phi},4_{\phi})&=&\frac{C_{F^2\phi^2}}{\Lambda^2}\la 12\ra ^2\,,
\label{F2phi2}\\
{\cal A}_{F\psi^2\phi}(1_{V_-},2_{\psi},3_{\psi},4_{\phi})&=&\frac{C_{F\psi^2\phi}}{\Lambda^2}\la 12\ra \la 13\ra\,, \label{dipole}\\
{\cal A}_{\psi^4}(1_{\psi},2_{\psi},3_{\psi},4_{\psi})&=&\left(C_{\psi^4}\la 12\ra \la 34\ra+C'_{\psi^4}\la 13\ra \la 24\ra\right)\frac{1}{\Lambda^2}\,.
\label{psi4}
\eea
With $h=0$, we have:
\bea
{\cal A}_{\square  \phi^4}(1_{\phi},2_{\phi},3_{\phi},4_{\phi})&=&
\left(C_{ \square \phi^4}\la 12\ra [12]+C'_{ \square \phi^4}\la 13\ra [13]\right)\frac{1}{\Lambda^2}\,,
\label{phi4}\\
{\cal A}_{\psi\bar\psi \phi^2}(1_{\psi},2_{\bar\psi},3_{\phi},4_{\phi})&=&\frac{C_{\psi\bar\psi \phi^2}}{\Lambda^2}\la 13\ra [23]\,,\\
{\cal A}_{\psi^2\bar\psi^2}(1_{\psi},2_{\psi},3_{\bar\psi},4_{\bar\psi})&=&\frac{C_{\psi^2\bar\psi^2}}{\Lambda^2}\la 12\ra [34]\,.
\label{psi2barpsi2}
\eea
\item 
{\bf n=5:}  On dimensional grounds,  these amplitudes must have one power of brackets (or squares). 
We have only one possibility, with $h=-1$:
\be
{\cal A}_{\psi^2\phi^3}(1_{\psi},2_{\psi},3_{\phi},4_{\phi},5_\phi)=\frac{C_{\psi^2\phi^3}}{\Lambda^2}\la 12\ra\,.
\ee

\item 
{\bf n=6:} This has dimension mass$^{-2}$, so it cannot carry any power of momentum. The only possibility is a 6-scalar amplitude, with $h=0$:
\be
{\cal A}_{\phi^6}(1_\phi,2_\phi,3_\phi,4_\phi,5_\phi,6_\phi)=\frac{C_{\phi^6}}{\Lambda^2}\,.
\label{last}
\ee
\end{itemize}
The corresponding complex-conjugated amplitudes
are obtained by the exchange $\la ij\ra\leftrightarrow [ji]$, and have opposite helicities, $h\to -h$. 
We notice that these  amplitudes can be unambiguously 
specified by assigning ($n$, $h$, $n_F$), where $n_F=0,2,4$ labels the fermion content.

As we said, the approach followed here is   equivalent to that with operators.
In fact,  if we choose a basis of higher-dimensional operators written in Weyl spinor notation
(see for instance \cite{Elias-Miro:2014eia}  for the case of the SM), 
the correspondence between dimension-6 operators and the above amplitudes is one-to-one.
For example, the amplitudes  of 
\eq{F3d}  and \eq{dipole}
correspond to the tree-level amplitudes with the lowest number of legs  that can be made, respectively, from the
dimension-6 operators
$F^{\alpha\beta}F_{\beta\gamma}F^{\gamma}_\alpha\equiv F^3$ 
 and  
$F^{\alpha\beta}\psi_{\alpha}\psi_{\beta}\phi\equiv F\psi^2\phi$, and  similarly for all the others.
In Appendix~\ref{app3} we give the explicit relation of some dimension-6 operators, written in the more usual Dirac notation \cite{Grzadkowski:2010es}, 
with  the on-shell amplitudes.
An advantage of  on-shell amplitudes versus operators is that we do not need to bother in specifying the operator basis, nor to eliminate redundancies by field redefinitions.

We will generically refer to the   amplitudes (\ref{F3d})--(\ref{last}) as ${\cal A}_{{\cal O}_i}$,
and their corresponding 
 coefficients as $C_{{\cal O}_i}$.
 These last play a similar   role as  the Wilson coefficients.
At the loop level, they can mix and lead to 
 an anomalous-dimension matrix equivalent to that in \eq{defAD}.
Below, we  discuss how to  calculate $\gamma_{i}$ using  unitarity methods.

\section{Anomalous dimensions from on-shell methods}
\label{sec2}

At the  one-loop level, any amplitude can have a Passarino-Veltman decomposition, given by
\be
{\cal A}_{\rm loop}=\sum_a C_2^{(a)} I_2^{(a)}+\sum_b C_3^{(b)} I_3^{(b)}+\sum_c C_4^{(c)} I_4^{(c)}+R\,,
\label{general}
\ee
where $I_m$  are master scalar integrals with $m$ propagators\footnote{Tadpole contributions cancel for massless theories, when using dimensional regularization.} ($m=2,3,4$) and $C_m$ are kinematic-dependent coefficients, rational functions of $\la ij\ra$ and $[ij]$. The master integrals are given by
\be
I_m=(-1)^m\mu^{4-D}\int\frac{d^D\ell}{i(2\pi)^D}\frac{1}{\ell^2(\ell-P_1)^2(\ell-P_1-P_2)^2\cdots}\,,
\ee
where  $P_1,P_2,...,P_{m-1}$ are sums of external momenta. 
We will be using dimensional regularization, $D=4-2\epsilon$, and always assume massless states.
The first three contributions to \eq{general} are called respectively bubbles, triangles and 
boxes, according to the topology of the scalar integral. 
Terms collected under $R$ are rational functions of the kinematical invariants. They will not play a relevant role in our analysis.

The expression \eq{general} is completely generic. Therefore it is perfectly suited to discuss universal properties of loop amplitudes.
The anomalous dimensions, in particular,
are related to the logarithmically UV divergent part of the amplitude. This means 
that they receive contributions only from  bubble integrals $I_2$,
since  $I_3$ and $I_4$ are  both UV convergent.
More explicitly, using dimensional regularization, we have
\be
I^{(a)}_2=\frac{1}{16\pi^2}\left( \frac{1}{\epsilon}+\ln\left(\frac{\mu^2}{-P_a^2}\right)+\cdots\right)\,,
\label{I2div}
\ee 
where $P_a$ is the sum of external 4-momenta that enters the bubble. 

These UV divergencies must be proportional to tree-level  amplitudes, due to the locality of the counterterms.
Here, we are interested in UV divergencies that appear at order $E^2/\Lambda^2$
and  renormalize the coefficients $C_{{\cal O}_i}$ discussed in the previous section.
We must then consider  one-loop amplitudes ${\cal A}_{\rm loop}$
with the same external legs as the amplitude that we want to renormalize, ${\cal A}_{{\cal O}_i}$,
and involving  one   (and only one) 
${\cal A}_{{\cal O}_j}$ in each loop.
In this case,   the sum of the UV divergencies  
of ${\cal A}_{\rm loop}$ is expected to  be proportional to  ${\cal A}_{{\cal O}_i}$:
\be
\frac{1}{8\pi^2}\sum_a C^{(a)}_2\propto {\cal A}_{{\cal O}_i}\,,
\label{gammasimp}
\ee
where we have used  \eq{general} and \eq{I2div}.
For the brevity  of the discussion, we are  only considering here the case where a unique
${\cal A}_{{\cal O}_i}$ appears on the RHS of \eq{gammasimp}.
We will come back to this point at the end of the section, where we discuss the more general situation.

One could  be tempted to associate the proportionality constant in \eq{gammasimp}
 to the anomalous dimension 
 $\gamma_i$ of the coefficient   $C_{{\cal O}_i}$.
Unfortunately, this is not so simple. 
To understand why, we must follow the fate of so-called ``massles'' bubbles, those for which $P^2_a=0$.

Massless bubbles do not contribute in \eq{general} because,  
for $P^2_a=0$, we have that  $I^{(a)}_2$
 is  dimensionless  and vanishes.
This  can be understood as an ``unwanted'' cancellation between UV and IR divergencies,
that  happens for terms proportional to $\ln\left(\mu_{\rm UV}/\mu_{\rm IR}\right)$, which vanish when using dimensional regularization where $\mu_{\rm UV}=\mu_{\rm IR}=\mu$.
Then, in order to obtain the full  contribution to $\gamma_i$, we  have to calculate separately the IR divergencies of the amplitude and subtract them off. IR divergencies are proportional to
the tree-level amplitude, and so
the  anomalous dimension can be expressed as
\be
\gamma_i \, {\cal A}_{{\cal O}_i}=-\frac{C_{{\cal O}_i}}{8\pi^2}\sum_a C^{(a)}_2+\gamma_{\rm IR}\, {\cal A}_{{\cal O}_i}\,.
\label{gamma}
\ee

 Fortunately, $\gamma_{\rm IR}$ is zero in many cases.
For instance, IR divergencies are absent when calculating the renormalization
of ${\cal A}_{{\cal O}_i}$
from another  amplitude ${\cal A}_{{\cal O}_j}$ with different number of legs, helicities or species.
 Also, they do not appear in  renormalizations that only involve  4-vertices, as can be the case for scalars (this is because massless
topologies are automatically absent in these theories).
In this article, we will consider  only those cases with $\gamma_{\rm IR}=0$.
We leave for a future work the  $\gamma_{\rm IR}\not=0$ case that includes, for example,
certain self-renormalization of the coefficients $C_{{\cal O}_i}$.

When IR divergencies are not present, we
can calculate the anomalous dimensions from only knowing the $C^{(a)}_2$ associated to ``massive'' bubbles. These bubble coefficients can 
be  obtained by using generalized unitarity methods, as described for instance 
in Refs.~\cite{Dixon:2013uaa,ArkaniHamed:2008gz}.
The coefficients $C_2^{(a)}$ are obtained by performing all possible double cuts (2-cuts)
of the loop amplitude, \eq{general}. A 2-cut is defined operationally through the Cutkosky rule
of putting two loop propagators on-shell, reducing $\mathcal{A}_{\rm loop}$ to a phase space integral
of two tree-level amplitudes. The most relevant property of 2-cuts is that
they are in one-to-one correspondence with the bubble coefficients. In other words,
each 2-cut picks up a unique $C_2^{(a)}$. The problem is that, in general,
2-cuts can also contain terms coming from triangles and boxes.

One way to disentangle $C^{(a)}_2$ from the rest is to first determine $C^{(c)}_4$ and $C^{(b)}_3$ by calculating quadruple 
and triple cuts, and then properly subtract them off from the 2-cut.
But this is  a lengthy procedure.

We will show below, however, that at the one-loop order and for amplitudes at order $1/\Lambda^2$,
the anomalous dimension of $C_{{\cal O}_i}$ can be simply obtained  as a sum over  2-cuts of the one-loop amplitude, giving
\be
\gamma_{ij}\,{\cal A}_{{\cal O}_i}(1,2,...,n)=-\frac{1}{4\pi^3}\frac{C_{{\cal O}_i}}{C_{{\cal O}_j}}
\int d{\rm LIPS}\sum_{\rm ext.\, legs\atop distrib.}\sum_{\ell_1,\ell_2}{\sigma_{\ell_1\ell_2}} \widehat {\cal A}_{{\cal O}_j}(...,\ell_1,\ell_2)\times {\cal A}_{4}(-\ell_2,-\ell_1,...)\,,
\label{MF}
\ee
{without summation over $i,j$}. Here, 
$\widehat {\cal A}_{{\cal O}_j}$ are $n\geq 4$ tree-level  amplitudes  containing  an
order $1/\Lambda^2$  amplitude ${\cal A}_{{\cal O}_j}$, that we classified in Eqs.~(\ref{F3d})--(\ref{last}), and
${\cal A}_{4}$ are   tree-level amplitudes made from marginal couplings  of the theory (dimension-4 operators), with $n\geq 4$.
{The dots in the arguments of $\widehat {\cal A}_{{\cal O}_j}$ and ${\cal A}_{4}$ stand for the external legs ($1,2,...,n$), that are distributed among the two amplitudes.\footnote{When the order of the external legs on the RHS of \eq{MF} differs from the one of the LHS, i.e. $1,2,...,n$, a minus sign must be included for each fermion exchanged.}  
A summation is included over the possible distributions of external legs (corresponding to the different 2-cuts).}
See Figs.~\ref{psi4todipole}--\ref{W3todipole}
for examples that we will be considering soon.
The  absence of  $n=3$ amplitudes  in \eq{MF} is   due to the fact that they can only lead to massless bubbles that, as we said,
vanish in dimensional regularization. This  fact  helps  in reducing the terms contributing 
to \eq{MF},    simplifying enormously the calculations. 

The integral  in \eq{MF} is over the   Lorentz-Invariant Phase Space (LIPS) associated with the two cut momenta, $\ell_1$ and $\ell_2$:
\be
 \int d{\rm LIPS}=  \int d^4\ell_1d^4\ell_2\, \delta^+(\ell^2_1)\delta^+(\ell^2_2)\delta^{(4)}(\ell_1+\ell_2-P)\,,
\label{dlips1}
 \ee
 where $P$ is the sum of the momentum of the external legs appearing in ${\cal A}_4$.  The integration measure is normalized as $ \int d{\rm LIPS}=\pi/2$,
 which is the reason why \eq{MF} carries an extra factor of $1/\pi$ besides the expected 
 $1/\pi^2$.
\eq{MF} also
includes a sum $\sum_{{\ell_1},{\ell_2}}$ over all possible internal states with momentum 
$\ell_1$ and $\ell_2$. 
 {The term $\sigma_{\ell_1\ell_2}$ is defined by $\sigma_{i_1i_2}\equiv i^{F[i_1i_2]}$, where $F[i_1i_2]$ counts the number of fermions in the list $\{i_1i_2\}$ (internal fermions). This factor arises from the convention in \eq{signs}, as explained in Appendix \ref{app2a}.}
The internal states in $ {\cal A}_4$ carry momentum, helicity and all other quantum numbers
with opposite sign with respect to those in $\widehat{\cal A}_{{\cal O}_j}$.
{For how to treat spinors with negative momenta, see also Appendix \ref{app2a}}.
A factor $1/2$ must be included when the internal particles are indistinguishable.

As we said, triangle and box contributions, that can be nonzero and pollute the 2-cuts, surprisingly cancel out in \eq{MF} at the order we are working.
In  Appendix~\ref{app1} we give a direct proof of this for the cases with $ n_i-n_j\equiv \Delta n< 2$.
We explicitly show how the cancellation of the loop
IR divergencies, which arise precisely from boxes and triangles, guarantees that their total contribution to 2-cuts is zero.

\begin{figure}[t]
\centering
\includegraphics[width=0.4\textwidth]{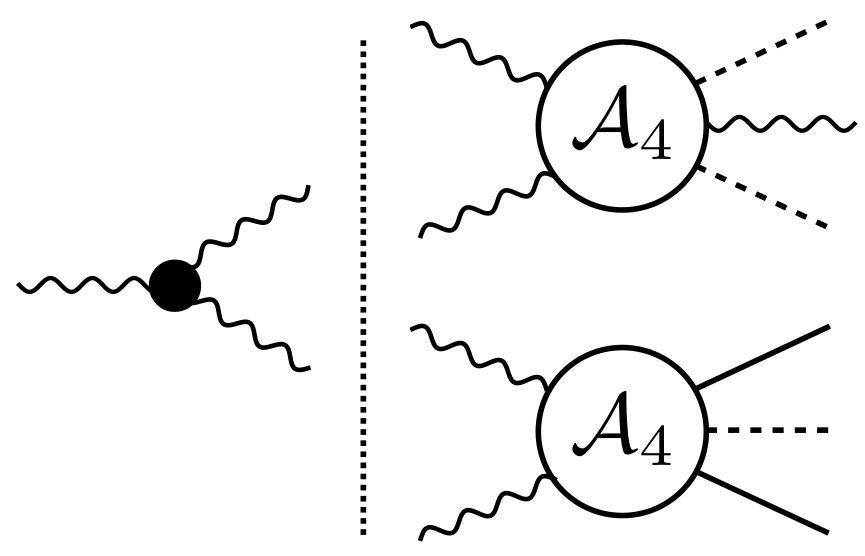}
\caption{\it 
Potential extra contributions  to the anomalous dimension of ${\cal F}_{F^2\phi^2}$
and ${\cal F}_{F\psi^2\phi}$  arising  from ${\cal F}_{F^3}$.}
\label{3to4}
\end{figure}

For a generic $\Delta n$,  the proof of \eq{MF} goes as follows.
In  \cite{Caron-Huot:2016cwu},
the following relation was derived, rewritten here for our particular case (see also \cite{Bern:2019wie}):
\be
\gamma_{ij}\, {\cal F}_{{\cal O}_i}(1,2,...,n)=
-\frac{1}{4\pi^3}
\int d{\rm LIPS}
\sum_{\rm ext.\, legs\atop distrib.}
 \sum_{{\ell_1},{\ell_2}} {\sigma_{\ell_1\ell_2}}
 \widehat {\cal F}_{{\cal O}_j} (...,\ell_1,\ell_2)
 \times
  {\cal A}_{4}(-\ell_2,-\ell_1,...) \,,
\label{CH}
\ee
where $\gamma_{ij}$ is the anomalous dimension matrix element of the form-factor ${\cal F}_{{\cal O}_i}$
associated to the dimension-6 operator ${\cal O}_i$:
\be
{\cal F}_{{\cal O}_i}(1,2,...,n)\equiv\la 0|{\cal O}_i| p_1,p_2, ...p_n\ra\,.
\ee
{The total momentum is not assumed here to be zero: $p_1+p_2+\cdots+p_n\equiv Q\not=0$.
By $\widehat {\cal F}_{{\cal O}_j}$ we again refer to form-factors containing the ``elementary'' form-factor  ${\cal F}_{{\cal O}_j}$.
Notice that $\widehat {\cal F}_{{\cal O}_j}$  can be a $n=3$ form-factor (as for example the contribution of Fig.~\ref{3to4}), since we have $Q\not=0$ and therefore these  contributions  are not 2-cuts of massless bubbles.
Now, taking  the limit  $Q\to 0$, we have
\be
\frac{C_{{\cal O}_i}}{\Lambda^2}{\cal F}_{{\cal O}_i}(1,2,...,n)\rightarrow
{\cal A}_{{\cal O}_i}(1,2,...,n)\,,
\label{limit}
\ee
and the terms in  \eq{CH}  must match to those of \eq{MF}, with the exception of the  terms in \eq{CH} containing $n=3$ form-factors. These latter,  in the limit  $Q\to 0$, lead to $n=3$ amplitudes that are absent in \eq{MF} as we already explained.\footnote{We remark that these  terms  can  only contain contributions from triangles or boxes, because  the terms in \eq{MF}, that arise from 2-cuts, already grasp all possible contributions from bubbles.}
For our particular case where ${\cal O}_i$ are dimension-6 operators,
it is easy to realize that the  only  contributions of this type  to \eq{CH} are those shown in  Fig.~\ref{3to4}.
We must show that these contributions are zero
in order to guarantee that  the limit  $Q\to 0$ brings \eq{CH}  to \eq{MF}.

The contributions of Fig.~\ref{3to4} correspond to the renormalizations of
${F^3 }$ to ${F^2\phi^2}$ and ${F\psi^2\phi}$, having both $\Delta n=1$.
But  for  $\Delta n=1$ contributions, we already   proved (with the use of  Appendix~\ref{app1})
the validity of \eq{MF}. Therefore
 the limit \eq{limit}  must indeed bring \eq{CH}  to \eq{MF}.
In other words,
 the  contributions of Fig.~\ref{3to4}
   must go to zero for $Q\to 0$.\footnote{This is not in general true, as can be seen from the examples in \cite{Caron-Huot:2016cwu}
where the anomalous dimension of
marginal operators is calculated.}
We have  checked this explicitly in the example of Section~\ref{W3}.
This completes the proof of  \eq{MF}.}

\begin{figure}[t]
\centering
\includegraphics[width=0.4\textwidth]{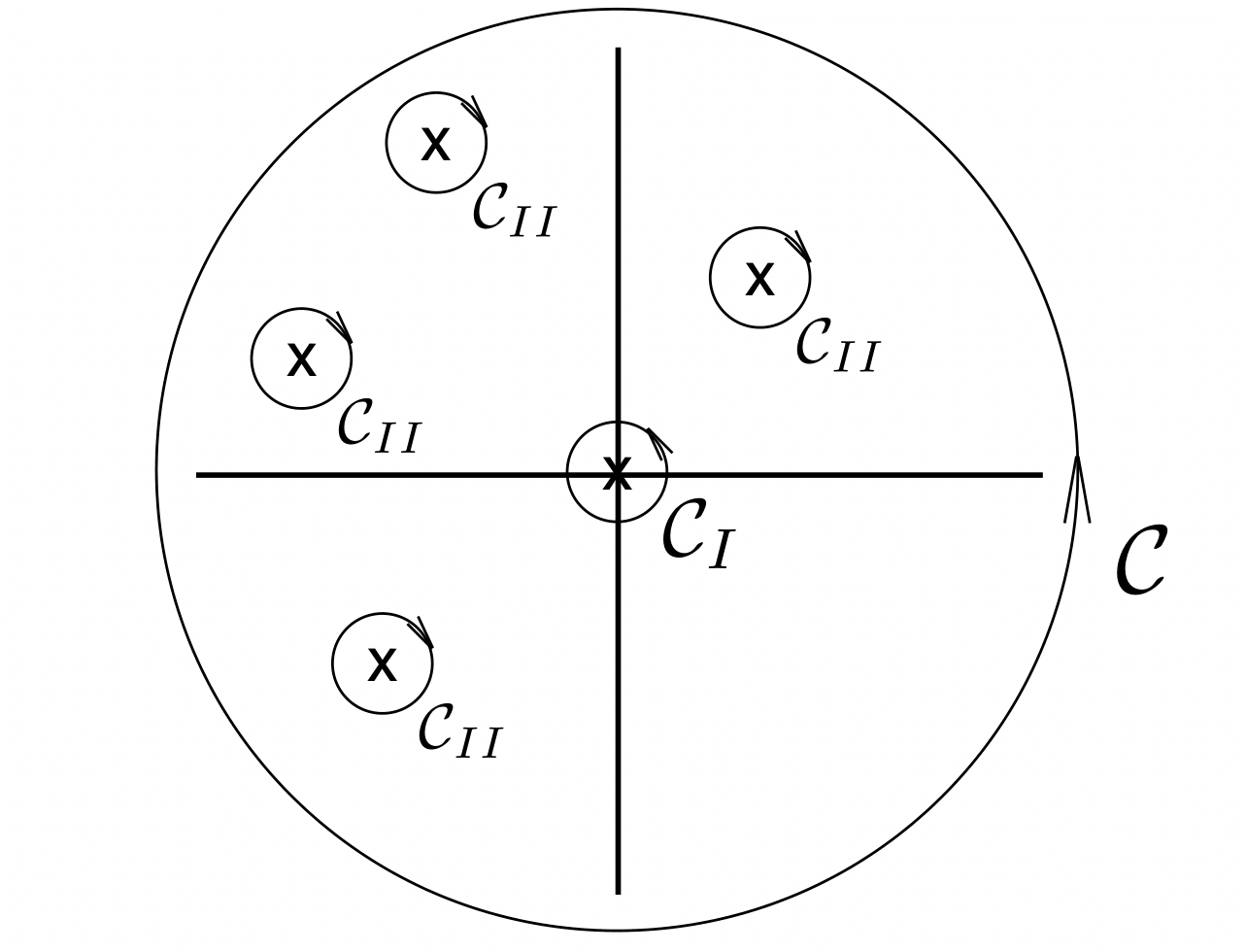}
\caption{\it Contours of integration in the complex $z$-plane. The contour ${\cal C}_I$ can be deformed to 
the contour ${\cal C}+{\cal C}_{II}$.}
\label{zplane}
\end{figure}

Let us also comment here on an alternative  method,  proposed in 
Refs.~\cite{ArkaniHamed:2008gz,Huang:2012aq},  to obtain each $C^{(a)}_2$ individually, using only 2-cuts (for other ways to extract bubble coefficients, see e.g. \cite{Forde:2007mi}).
This is based on a BCFW deformation \cite{Britto:2005fq} of the cut legs, sending $\ell_1\to  \ell_1 + qz$ and  $\ell_2\to  \ell_2-qz$,
that promotes the integrand of \eq{MF} to a complex function of $z$. Using the standard `Cauchy trick', we can
rewrite the integrand as a contour integral in $z$ (see Fig.~\ref{zplane}) along contour ${\cal C}_I$:
\be
\widehat {\cal A}_{{\cal O}_j}(...,\ell_1,\ell_2)\times {\cal A}_{4}(-\ell_2,-\ell_1,...)=
\frac{1}{2\pi i}
\int_{{\cal C}_I} \frac{dz}{z}
 \widehat {\cal A}_{{\cal O}_j}(...,\ell_1(z),\ell_2(z))\times {\cal A}_{4}(-\ell_2(z),-\ell_1(z),...)\,.
 \ee
The complex integrand is a product of tree amplitudes. Because of this, its singularities can only be poles coming from
propagators going on-shell. By deforming  the contour ${\cal C}_I$ as in Fig.~\ref{zplane},  we have
$
\int_{{\cal C}_I} dz=\int_{{\cal C}} dz+\int_{{\cal C}_{II}} dz
$.
The poles that are picked up by ${\cal C}_{II}$ must be associated to triangles and boxes, since they are  the only scalar diagrams that remain with
uncut propagators (after the 2-cut). 
If we drop these, we are left with the integral over ${\cal C}$, that selects
the pole at infinity. {As explained in \cite{ArkaniHamed:2008gz}}, this is precisely due to the presence of bubbles.
We then have \cite{ArkaniHamed:2008gz}
\be
\gamma_{ij}{\cal A}_{{\cal O}_i}(1,2,...,n)\hspace{-0.2em}=
i\frac{C_{{\cal O}_i}}{C_{{\cal O}_j}}\hspace{-0.2em}\int \hspace{-0.2em} \frac{d{\rm LIPS}}{8\pi^4}\hspace{-0.4em}\sum_{\rm ext.\, legs\atop distrib.}\hspace{-0.1em}\sum_{{\ell_1},{\ell_2}}\hspace{-0.2em}{\sigma_{\ell_1\ell_2}}\hspace{-0.4em}\int_{{\cal C}}\hspace{-0.2em} \frac{dz}{z}
 \widehat {\cal A}_{{\cal O}_j}\hspace{-0.1em}(...,\ell_1(z),\ell_2(z)\hspace{-0.05em})\hspace{-0.2em}\times \hspace{-0.2em}{\cal A}_{4}(\hspace{-0.2em}-\ell_2(z),\hspace{-0.2em}-\ell_1(z),\hspace{-0.1em}...)\,.
\label{ACK}
\ee
The integral over ${\cal C}$  can be  equivalently obtained by  extracting the constant term in a Laurent series around $\infty$ of the $z$-dependent product of  amplitudes.  
Although \eq{ACK} looks more  involved than \eq{MF}, in those cases in which 
contributions from boxes are nonzero in the individual 2-cuts,
the calculation of the anomalous dimension from \eq{ACK} is in practice much easier.
While for renormalizations with $\Delta n=0$ 
triangle and box contributions are not present (see Appendix \ref{app1}),
and then it is pointless to use \eq{ACK},
for $\Delta n\geq 1$ processes, instead, triangles and boxes can appear,
and it turns very useful to project them out with \eq{ACK}.
We will see an explicit example in Section~\ref{W3}.
 
 We close this chapter with a few additional remarks.
Although the derivation of  \eq{MF} came  from performing 2-cuts of one-loop Feynman diagrams, 
we do not need to  refer anymore to   loop diagrams when calculating anomalous dimensions. Indeed,
\eq{MF} tells us  that we just need to sum over   all possible products of two $n\geq 4$ tree-level amplitudes, one made with ${\cal A}_{{\cal O}_j}$ and the other with SM vertices,   
with the following conditions satisfied: ($i$) the two amplitudes must share two legs (identical up to a conjugation), the so-called internal legs, 
($ii$) the rest of their legs (the external ones)  must match  those of  ${\cal A}_{{\cal O}_i}$.
We will see many explicit examples in the next section.

Another thing worth mentioning about  \eq{MF} 
are the following obvious rules that it fulfills:
\bea
  n_i&=&\widehat n_{j}+n_{4}-4\,,
  \label{nsr}
\\
  h_i&=&\widehat h_{j}+h_{4}\,,
  \label{hsr}
\eea
where $n_i$ ($\widehat n_{ j}$)
is the number of legs of ${\cal A}_{{\cal O}_i}$  ($\widehat {\cal A}_{{\cal O}_j}$) and 
$n_4$ the number of legs of ${\cal A}_4$, and similarly for the helicities.
Since $\widehat n_{j}\geq n_j$ and $n_4\geq 4$, we derive from \eq{nsr}:
\be
\Delta n\geq 0\,,
\label{deltanequal0}
\ee
that tells us that   $C_{{\cal O}_j}$ can  contribute to the
 anomalous dimensions  of $C_{{\cal O}_i}$ only if ${\cal A}_{{\cal O}_j}$
 has equal or less number of legs than ${\cal A}_{{\cal O}_i}$ (see 
Ref.~\cite{Bern:2019wie} for an extension of this rule to higher loop orders).
Furthermore,  since almost all  $n=4$  amplitudes  made from marginal couplings   have $h=0$,
 we have that ${\cal A}_4$, when  built from these amplitudes, will  fulfill  $n_4\geq |h_4|+4$. This allows to   derive, together with Eqs.~(\ref{nsr})--(\ref{hsr})  the  selection rule \cite{Cheung:2015aba}\footnote{Selection rules can also be derived
 using supersymmetry \cite{Elias-Miro:2014eia} or 
 angular momentum conservation \cite{Jiang:2020sdh}.
See also Ref.~\cite{Craig:2019wmo} for an alternative derivation.} 
\be
 \Delta n\geq |\Delta h|\,.
\label{sr}
\ee
The only  exceptions to \eq{sr} come from one-loop amplitudes
involving the only $n=4$  ${\cal A}_4$   that has $|h|> 0$: this is the  4-fermion  $\psi^4$ amplitude, 
that has $h=-2$ and can for example be generated in the SM
by the exchange of the Higgs.
Nevertheless,  one-loop contributions from  $\psi^4$ can only violate the rule 
\eq{sr} in the renormalization  
between amplitudes with very specific properties, fulfilling  $\Delta n_F=0$ and   $|\Delta h|=2$. That is,
only between $C_{\psi^4}$ and $C_{\psi^2\bar\psi^2}$ 
or between
$C_{H^3\psi^2}$ and $C_{H^3\bar\psi^2}$.
We will see  applications of the above selection rules  in the next section.

There is also another very useful  selection rule which will  allow us to derive  new non-renormalization theorems.
 As \eq{MF} shows, symmetries of the external legs 
of  $\widehat {\cal A}_{{\cal O}_j}$ or  ${\cal A}_{4}$
must also be symmetries  of the renormalized ${\cal A}_{{\cal O}_i}$. 
 This is of course true whenever the  symmetry
property is shared by all the contributions to a given renormalization.
This implies, as we will see in the next examples, that not only global symmetries, but also (anti)symmetries under the exchange of  external spinors  can lead to interesting
non-renormalization properties.

Up to now,  we have considered  \eq{MF}  for the cases in which, 
for a given amplitude, which is determined by the external states, there is a unique 
${\cal A}_{{\cal O}_i}$  contributing at tree-level. 
Nevertheless, there are certain  cases where there can be more than one 
${\cal A}_{{\cal O}_i}$  contributing, 
and \eq{MF} must be modified. 
These  cases are 
\begin{itemize} 
\item  
${\cal A}(1_{V_-},2_{V_-},3_{\phi},4_{\phi})$ where ${\cal A}_{F^2\phi^2}$ contributes 
as a contact-interaction, but also ${\cal A}_{F^3}$
as a sub-amplitude, with one of the  $V_-$  propagating   to end up in   a  $\phi\phi^\dagger$.
\item 
${\cal A}(1_{\psi},2_{\psi},3_{\phi},4_{\phi},5_\phi)$
where ${\cal A}_{\psi^2\phi^3}$ contributes 
as a contact-interaction, but also
{${\cal A}_{F\psi^2\phi}$, ${\cal A}_{\psi\bar{\psi}\phi^2}$} and ${\cal A}_{\square \phi^4}$ can enter 
as sub-amplitudes.
\item 
${\cal A}(1_\phi,2_\phi,3_\phi,4_\phi,5_\phi,6_\phi)$ where ${\cal A}_{\phi^6}$ contributes 
as a contact-interaction, 
but also ${\cal A}_{F^3}$, ${\cal A}_{F^2\phi^2}$ and  ${\cal A}_{\square \phi^4}$ as sub-amplitudes.
\end{itemize}
Since each ${\cal A}_{{\cal O}_i}$ enters with different 
 $\la ij\ra$ and $[ij]$ dependence in the corresponding amplitude,
  we can easily disentangle 
 the contributions   to the anomalous dimension of each of them.
 For example, the contributions to ${\cal A}(1_{V_-},2_{V_-},3_{\phi},4_{\phi})$
 from  ${\cal A}_{F^2\phi^2}$ and  ${\cal A}_{F^3}$  are respectively given by \eq{F2phi2}
and \eq{F3toF2h2} below.
When calculating the RHS of \eq{MF} for this amplitude,
we will get some terms proportional
to \eq{F2phi2},
and some others to \eq{F3toF2h2}. Only the first ones correspond to the anomalous dimension
of $C_{F^2\phi^2}$. 
We will present  examples  of this type in a future work.

Similarly, 
when different ``flavors" are added,  like in the SM,
there can be several independent coefficients $C_{{\cal O}_i}$  
contributing to the same process.
Nevertheless, by projecting \eq{MF} on a basis of invariant tensors under Lorentz and the global symmetries,  it is easy to identify the anomalous contribution to each $C_{{\cal O}_i}$.
We will see an example in the next section.

We  finally  mention that \eq{MF} can also be applied to find the anomalous dimensions of 
 dimension-5 operators.
This is  shown  in Appendix~\ref{app4}.

\section{One-loop anomalous dimensions of the  SM dipole \\ operators}
\label{smcal}

As an example of the use, reach and simplicity  of \eq{MF}, we present in this section
the  calculation of all one-loop  anomalous dimensions of the  SU(2)$_L$ dipole operator of the electron  (up to self-renormalization).
This is equivalent to calculate 
 the anomalous dimension of the  coefficient $C_{F\psi^2\phi}$, defined  in \eq{dipole},
 for the particular case of  the SM.

\begin{figure}[t]
\centering
\includegraphics[width=0.25\textwidth]{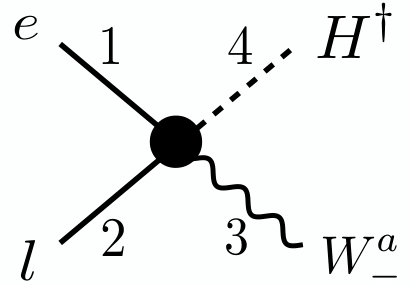}
\caption{\it Tree-level contribution to the $W_-^a  H^\dagger le$ amplitude.}
\label{dipoleSM}
\end{figure}

The amplitude  to consider is  $W_-^a H^\dagger  le $,
where $W^a_-$ is an SU(2)$_L$ gauge boson with $h=-1$,
 $H$ is the Higgs of hypercharge $Y_H=1/2$, 
and   $l$, $e$  are  respectively the SM SU(2)$_L$-doublet  and singlet 
leptons, with $h=-1/2$ and  hypercharges $Y_l=-1/2$ and $Y_e=1$.
At  tree-level, following the notation of Fig.~\ref{dipoleSM},
the only contribution to this amplitude is  given by
\be
{\cal A}(1_{e},2_{l_j},3_{W^a_-},4_{H^\dagger_i})=\frac{C_{WHle}}{\Lambda^2}\la 31\ra\la 32\ra (T^a)_{ij}\equiv {\cal A}_{WHle}\,,
\label{treelevel}
\ee
with $T^a=\sigma^a/2$ here. 
We recall that, for amplitudes involving fermions, respecting the order of labels is crucial for getting the signs correct (see Appendix~\ref{app2a} and references therein).
At the loop level, the coefficient $C_{WHle}$ receives an anomalous dimension,
that we will  denote by $\gamma_{WHle}$. 
Using \eq{sr} we can easily see that only a few  $C_{{\cal O}_i}$ can contribute to this anomalous dimension. Indeed, since \eq{treelevel} has $n=4$ and $h=-2$, only ${\cal A}_{{\cal O}_j}$  with $n=3$ or $n=4$, $h=-2$   can contribute. 
This leaves only  the coefficients of 
\eq{F3d} and Eqs.~(\ref{F2phi2})--(\ref{psi4})
as potential candidates to  contribute to the anomalous
dimension of $C_{WHle}$. 
We already see the usefulness of the amplitude method approach, allowing here
to easily understand that there are many vanishing contributions to the dipole operators.
In working within the usual Feynman diagram approach, these zeros   appear  as mysterious cancellations between different one-loop diagrams.

 We also notice that \eq{treelevel} is symmetric under the interchange of spinors
$1$ and $2$. As we will see, this property also  provides useful selection rules 
for non-renormalizations, 
that are often not apparent when using higher-dimensional operators in Dirac notation
\cite{Grzadkowski:2010es}.

\subsection{One-loop contribution   from $C_{\psi^4}$, $C_{F^2\phi^2}$ and $C_{F\psi^2\phi}$}
\label{dipoleren}

Let us start with the contributions from $n=4$ ${\cal A}_{{\cal O}_j}$ amplitudes.
We first consider ${\cal A}_{\psi^4}$.
We require   at least two SM leptons in order to contribute to $W^a_-H^\dagger le$. 
This leaves,
as the only possible set of negative-helicity fermions forming a SM singlet,
 the set  $e,l,q,u$, where $q$ and $u$ 
are  respectively the SM SU(2)$_L$-doublet  and singlet  quark, with $h=-1/2$ and  hypercharges $Y_q=-1/6$ and $Y_u=2/3$.
We have  then two possible   amplitudes\footnote{A third possibility $\propto\la 13\ra \la 42\ra$ can be reduced to the given  ones by the Schouten identity, \eq{Schouten}.}
\be
{\cal A}_{luqe}(1_{e},2_{l_i},3_{u},4_{q_j})=\frac{C_{luqe}}{\Lambda^2}\la 23\ra \la 41\ra
 \epsilon_{ij}
\,,
\label{luqe}
\ee
and 
\be
{\cal A}_{lequ}(1_{e},2_{l_i},3_{u},4_{q_j})=\frac{C_{lequ}}{\Lambda^2}\la 12\ra \la 34\ra \epsilon_{ij}\,.
\label{lequ}
\ee
Since \eq{lequ} is antisymmetric under $1\leftrightarrow 2$, it cannot contribute to \eq{treelevel}, that is symmetric.
We are then left with only \eq{luqe}.

\begin{figure}[t]
\centering
\includegraphics[width=0.5\textwidth]{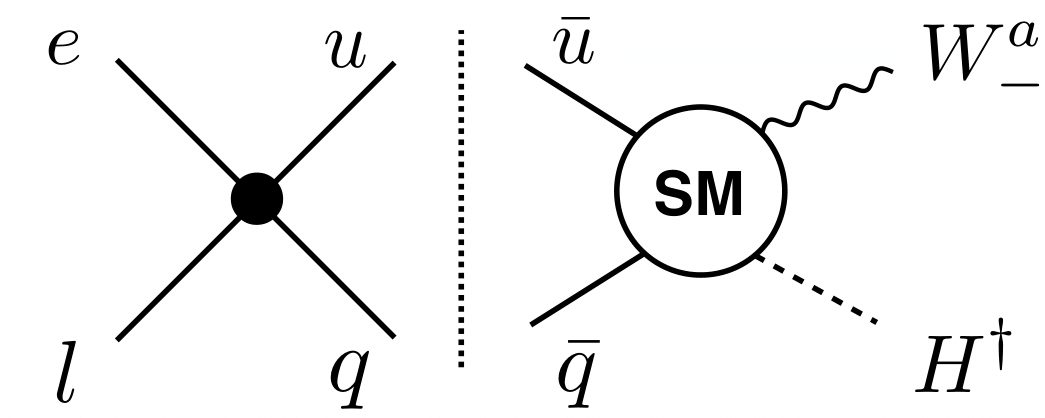}
\caption{\it Contribution from $C_{luqe}$ to the anomalous dimension of $C_{WHle}$.}
\label{psi4todipole}
\end{figure}

Following \eq{MF}, we can easily  calculate the contribution to the anomalous dimension of $C_{WHle}$ arising  from $C_{luqe}$.
We find that the  only possible contribution is the one that is diagrammatically pictured in Fig.~\ref{psi4todipole},
which gives (from now on we drop the $i,j$ SU(2)$_L$ indices)
\bea
\gamma_{WHle} \frac{\la 31\ra\la 32\ra\, T^a}{\Lambda^2}
&=&\frac{1}{4\pi^3}\int d{\rm LIPS}\, {\cal A}_{luqe}(1_e,2_{l},3'_u,4'_{q})\times {\cal A}_{\rm SM}(-4'_{\bar q},-3'_{\bar u},
3_{W^a_-},4_{H^\dagger})\nonumber\\
&=& -\frac{y_u g_2N_c}{4\pi^3}C_{luqe}\, T^a \int d{\rm LIPS}\, \frac{\la 23'\ra\la 4'1\ra}{\Lambda^2}\times \frac{\la 34\ra \la 33' \ra}{\la 43'\ra\la  3'4'\ra}
\,,
\label{fpsi2h}
\eea
where $N_c=3$, the $d$LIPS integration is taken over the primed spinors
with $p_{3'}+p_{4'}=p_3+p_4$, and we have used 
\eq{luqe} and \eq{su2}. 
A very convenient way to simplify this integral is to
 relate  the spinors $| 3'\ra$ and $| 4'\ra$
with the external spinors  $|3\ra$ and $|4\ra$, as explained in Ref.~\cite{Caron-Huot:2016cwu}:
\bea
|3'\ra&=&c_\theta|3\ra -s_\theta e^{i\phi}|4\ra\,,\nonumber\\
|4'\ra&=&s_\theta e^{-i\phi}|3\ra+c_\theta|4\ra\,,
\label{34primed}
\eea
where $s_\theta\equiv\sin\theta$ and $c_\theta\equiv\cos\theta$.
By complex conjugating  \eq{34primed},
we can get similar relations for  $|3']$ and $|4']$, and easily show that  $p_{3'}+p_{4'}=p_3+p_4$, identically for any $(\theta,\phi)$.
Using \eq{34primed}, the $d$LIPS integration is simplified to a solid angle integration  \cite{Caron-Huot:2016cwu}: 
\be
\frac{2}{\pi}\int d{\rm LIPS}\equiv\int^{2\pi}_0\frac{d\phi}{2\pi}\int^{\pi/2}_0d\theta\, 2 s_\theta c_\theta\,.
\label{dlips}
\ee
The integration over the angle $\phi$
projects the RHS of \eq{fpsi2h} into $\la 31\ra\la 32\ra$, leading to
\be
\gamma_{WHle}=\frac{y_ug_2N_c}{4\pi^2}C_{luqe} \int^{\pi/2}_0 d\theta\,  s^3_\theta c_\theta=\frac{ y_u g_2N_c}{16\pi^2}C_{luqe}\,.
\ee
It is important to notice that we did not have to use momentum conservation 
 in the on-shell amplitude ${\cal A}_{luqe}$.
Therefore, our calculation would have proceeded in the same way, if we had 
used \eq{CH} with  $p_1+p_2+p_{3'}+p_{4'}=Q\not=0$, 
taking the limit $Q\to  0$ at the end of the calculation.
This provides a check that \eq{CH} and \eq{MF} agree at this order.

\begin{figure}[t]
\centering
\includegraphics[width=0.5\textwidth]{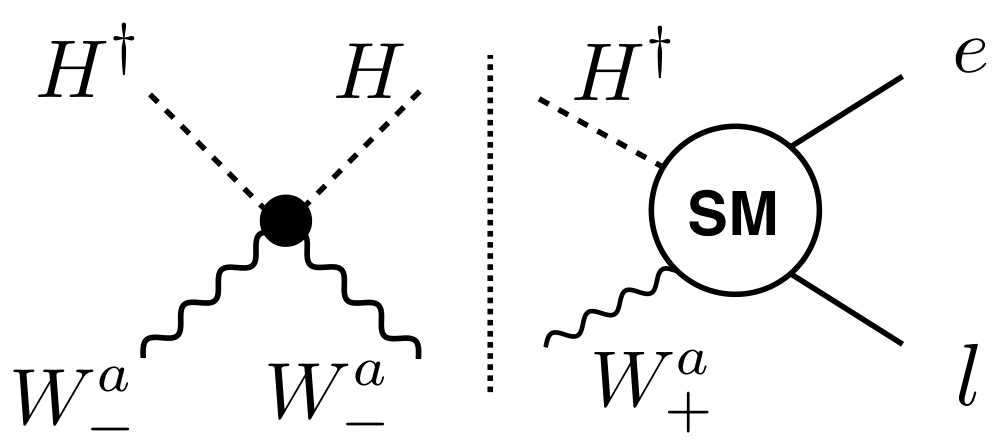}
\caption{\it Contribution from $C_{W^2H^2}$  to the anomalous dimension of $C_{WHle}$.}
\label{F2h2todipole}
\end{figure}

In the same simple way,  we can proceed with  the contribution from coefficients of type $C_{F^2\phi^2}$.
The contribution from an internal $W$ is shown diagrammatically in  Fig.~\ref{F2h2todipole}, and gives
\bea
\gamma_{WHle} 
&=&
-\frac{\Lambda^2}{\la 31\ra\la 32\ra  T^a}
\frac{1}{4\pi^3}\int d{\rm LIPS}\, {\cal A}_{W^2H^2}(3_{W_-^a},4_{H^\dagger},1'_{W_-^a},2'_{H})\times {\cal A}_{\rm SM}(-1'_{W^a_+},-2'_{H^\dagger},1_e,2_{l})\nonumber\\
&= &\frac{y_eg_2}{4\pi^3}\frac{C_{W^2H^2}}{\la 31\ra\la 32\ra}\int d{\rm LIPS}\, {\la 31'\ra^2}\times \frac{\la 2'2\ra \la 12 \ra}{\la 1'2'\ra\la  1'2\ra}\nonumber\\
&=& -\frac{y_eg_2}{2\pi^2}C_{W^2H^2}\int^{\pi/2}_0 \!\!\!\!d\theta\,  s^3_\theta c_\theta
=-\frac{y_eg_2}{8\pi^2}C_{W^2H^2}
\,,\label{fpsi2h2}
\eea
where we have used \eq{34primed}, adapted for relating  $|1'\ra$ and $|2'\ra$ with $|1\ra$ and $|2\ra$.
Similarly to \eq{fpsi2h2}, we have, for the case of an internal $B$:
\bea
\gamma_{WHle} 
&=&
-\frac{\Lambda^2}{\la 31\ra\la 32\ra  T^a}
\frac{1}{4\pi^3}\int d{\rm LIPS}\, {\cal A}_{WBH^2}(3_{W^a_-},4_{H^\dagger},1'_{B_-},2'_{H})\times {\cal A}_{\rm SM}(-1'_{B_+},-2'_{H^\dagger},1_e,2_{l})\nonumber\\
&=& \frac{y_eg_1}{4\pi^3}\frac{C_{WBH^2}}{\la 31\ra\la 32\ra}\int d{\rm LIPS}\, {\la 31'\ra^2}\times \left(Y_{l}\frac{\la 2'2\ra \la 12 \ra}{\la 1'2'\ra\la  1'2\ra}-Y_{e}\frac{\la 2'1\ra \la 21 \ra}{\la 1'2'\ra\la  1'1\ra}\right)\nonumber \\
&=& -\frac{y_eg_1}{2\pi^2}C_{WBH^2}\int^{\pi/2}_0 \!\!\!\!d\theta\,  \left(Y_l s^3_\theta c_\theta - Y_e s_\theta c^3_\theta \right)
=-\frac{y_eg_1}{8\pi^2}\left(Y_l-Y_e\right)C_{WBH^2}
\,.\label{fpsi2h2b}
\eea

At this point, it is worth noticing several interesting features
of this procedure.
First, we can see how the  two contributions of Fig.~\ref{psi4todipole} and \ref{F2h2todipole},
that from the Feynman diagrammatic viewpoint look so different, 
are very similar in the on-shell amplitude method, \eq{fpsi2h} and \eq{fpsi2h2},
due to similar helicity structure.
This universality in one-loop corrections    helps to avoid mistakes.
Furthermore,  once one is  armed with  the SM amplitude ${\cal A}_{\rm SM}(1_{V^a_+},2_{H^\dagger},3_\psi,4_{\psi})$,
one can easily calculate all $\gamma_{ij}$ non-diagonal terms
 between the different  $h=-2$ amplitudes, those of Eqs.~(\ref{F2phi2})--(\ref{psi4}). 
 This is because  we can go from one to the other by just 
 multiplying them  with the same amplitude  ${\cal A}_{\rm SM}(1_{V^a_+},2_{H^\dagger},3_\psi,4_{\psi})$,  
but  taking  different sets of internal legs in each case. 
This is an example of the  ``recycling'' power of the   on-shell method, in which new calculations 
nurture from previous ones,  without the need  of starting from scratch, as it 
is usually the case  in  the  Feynman diagram approach.
Another  example is  the one-loop mixing  between the amplitudes of Eqs.~(\ref{phi4})--(\ref{psi2barpsi2}), that  can be calculated from the same  SM amplitude: $HH^\dagger \psi\bar\psi$.

\begin{figure}[t]
\centering
\includegraphics[width=0.5\textwidth]{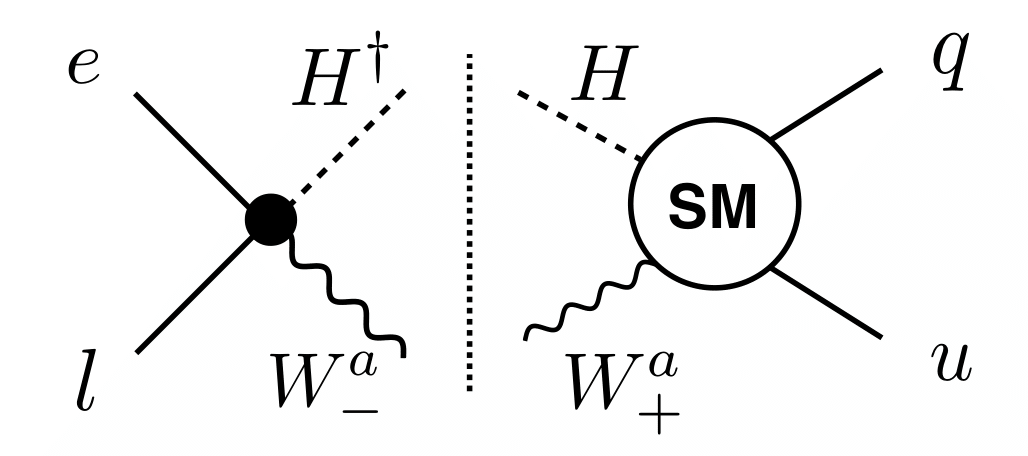}
\caption{\it Contribution from $C_{WHle}$  to the anomalous dimension of  $C_{luqe}$ and $C_{lequ}$.}
\label{dipoletopsi4}
\end{figure}

As an illustration of this recycling aspect,
we  consider here the ``inverse" of \eq{fpsi2h}, that is
the contribution of the dipole coefficient $C_{WHle}$ to 4-fermion amplitudes, \eq{luqe} and 
\eq{lequ}. 
The contribution is shown in Fig.~\ref{dipoletopsi4}, and it gives
\bea
\gamma_{lequ}\frac{{\cal A}_{lequ}}{C_{lequ}}+\gamma_{luqe} \frac{{\cal A}_{luqe}}{C_{luqe}}
&=&
-\frac{1}{4\pi^3}\int d{\rm LIPS}\, {\cal A}_{WHle}(1_{e},2_{l},3'_{W_-^a},4'_{H^\dagger})\times {\cal A}_{\rm SM}(-3'_{W^a_+},-4'_{H},3_u,4_{q})\nonumber\\
&= &\frac{y_u g_2}{4\pi^3}\frac{C_{WHle}}{\Lambda^2}(T^a)^2 \int d{\rm LIPS}\, {\la 3'1\ra \la 3'2\ra}\times \frac{\la 34\ra \la 4'4 \ra}{\la 3'4\ra\la  3'4'\ra}\nonumber\\
&=& -\frac{3y_u g_2}{64\pi^2}\frac{C_{WHle}}{\Lambda^2} \left(\la   31\ra \la 42\ra + \la  32\ra\la 41\ra\right)
\,,\label{ADpsi4}
\eea
where we have used  $(T^a)^2=3/4$. 
Notice that the fact that  \eq{treelevel} is symmetric under  $1\leftrightarrow 2$
 assures the form of \eq{ADpsi4}, i.e. it can only renormalize a combination
that is symmetric  under  $1\leftrightarrow 2$.
This  selection rule is  non-trivial from Feynman diagrams, since there are in principle loops 
in which the leptons of the dipole operator are in the internal lines.
Using the Schouten identity to project \eq{ADpsi4} into the 4-fermion amplitudes \eq{luqe} and \eq{lequ},
we obtain
\be
\gamma_{luqe}=-2\gamma_{lequ}=\frac{3y_u g_2}{32\pi^2}C_{WHle}\,.
\label{glequ}
\ee

\begin{figure}[t]
\centering
\includegraphics[width=0.6\textwidth]{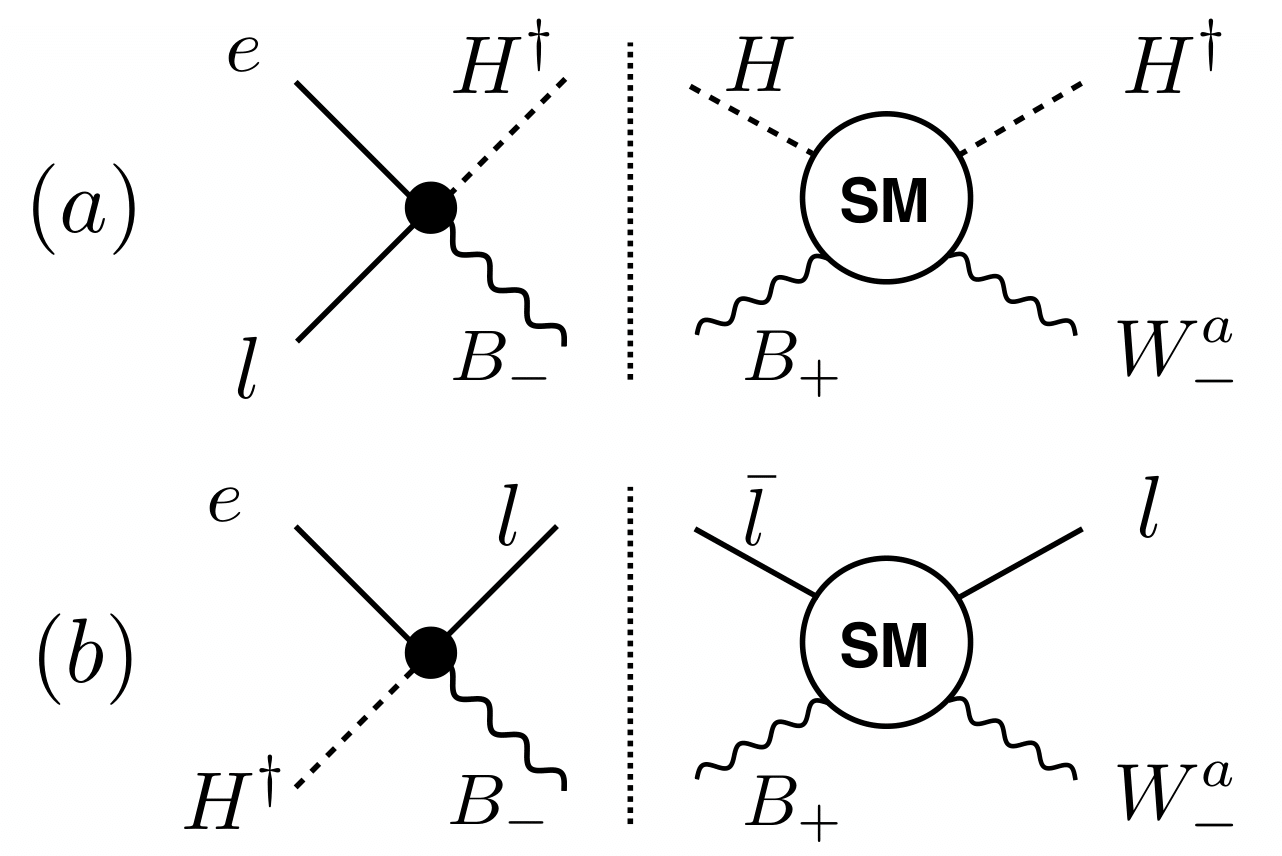}
\caption{\it Contributions   from $C_{BHle}$ to the anomalous dimension of $C_{WHle}$.}
\label{dipoletodipole}
\end{figure}

Finally, for completeness,   we also show the  calculation of the only  contribution to $\gamma_{WHle} $ coming from another  dipole operator, that  involving a  $B$. 
There are two contributions, as shown in Fig.~\ref{dipoletodipole}. 
The contribution from  $(a)$ gives
\bea
\gamma_{WHle} 
&=&
-\frac{\Lambda^2}{\la 31\ra\la 32\ra  T^a}
\frac{1}{4\pi^3}\int d{\rm LIPS}\, {\cal A}_{BHle}(1_{e},2_{l},3'_{B_-},4'_{H^\dagger})\times {\cal A}_{\rm SM}(-3'_{B_+},-4'_{H},3_{W^a_-},4_{H^\dagger})\nonumber\\
&= &\frac{g_1g_2Y_H}{4\pi^3}\frac{C_{BHle}}{\la 31\ra\la 32\ra}\int d{\rm LIPS}\, {\la 3'1\ra \la 3'2\ra}\times \frac{\la 4'3\ra \la 43 \ra}{\la 4'3'\ra\la  43'\ra}\nonumber\\
&=&  \frac{g_1g_2Y_H}{4\pi^2}C_{BHle}\int^{\pi/2}_0 \!\!\!\!d\theta\,  c^3_\theta s_\theta= \frac{g_1g_2}{16\pi^2}Y_H C_{BHle}\,,
\label{fpsi2h2c}
\eea
where we have used \eq{WBh2}.  The  contribution from $(b)$ of Fig.~\ref{dipoletodipole} gives
\bea
\gamma_{WHle} 
& =&
 -\frac{\Lambda^2}{\la 31\ra\la 32\ra  T^a}
\frac{{ i}}{4\pi^3}\int d{\rm LIPS}\, {\cal A}_{BHle}(1_{e},2'_{l},3'_{B_-},4_{H^\dagger})\times {\cal A}_{\rm SM}(-3'_{B_+},-2'_{\bar{l}},3_{W^a_-},2_{l})\nonumber\\
& = &  -\frac{g_1g_2Y_l}{4\pi^3}\frac{C_{BHle}}{\la 31\ra\la 32\ra}\int d{\rm LIPS}\, {\la 3'1\ra \la 3'2'\ra}\times \frac{\la 2 3 \ra^2}{\la 2 3'\ra\la  3'2'\ra}\nonumber\\
& =& \frac{g_1g_2Y_l}{4\pi^2}C_{BHle}\int^{\pi/2}_0 \!\!\!\!d\theta\,  s_\theta c_\theta=\frac{g_1g_2}{8\pi^2}Y_l C_{BHle}\,,
\label{fpsi2h2d}
\eea
 where we have used \eq{WBl2}. Taking into account that $Y_H=Y_l+Y_e$, the total contribution 
 from \eq{fpsi2h2c} and \eq{fpsi2h2d} gives
  \bea
\gamma_{WHle}=\frac{g_1g_2}{16\pi^2}\left(3Y_l +Y_e \right)C_{BHle}\,.
\label{fpsi2h2e}
\eea

\subsection{One-loop contribution from $C_{F^3}$}
\label{W3}

\begin{figure}[t]
\centering
\includegraphics[width=0.6\textwidth]{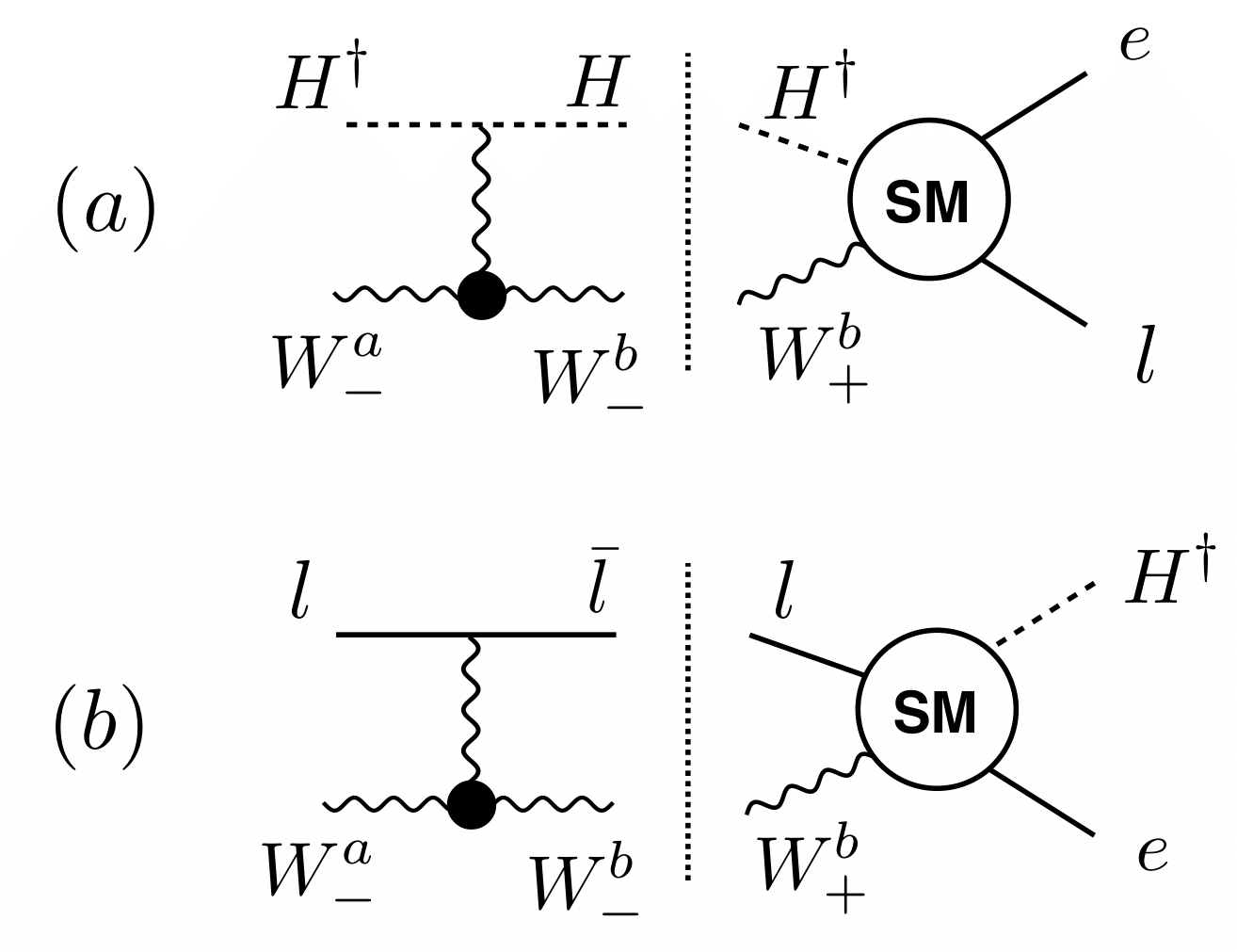}
\caption{\it Contributions from $C_{W^3}$ to the anomalous dimension of $C_{WHle}$.}
\label{W3todipole}
\end{figure}

\begin{figure}[t]
\centering
\includegraphics[width=0.5\textwidth]{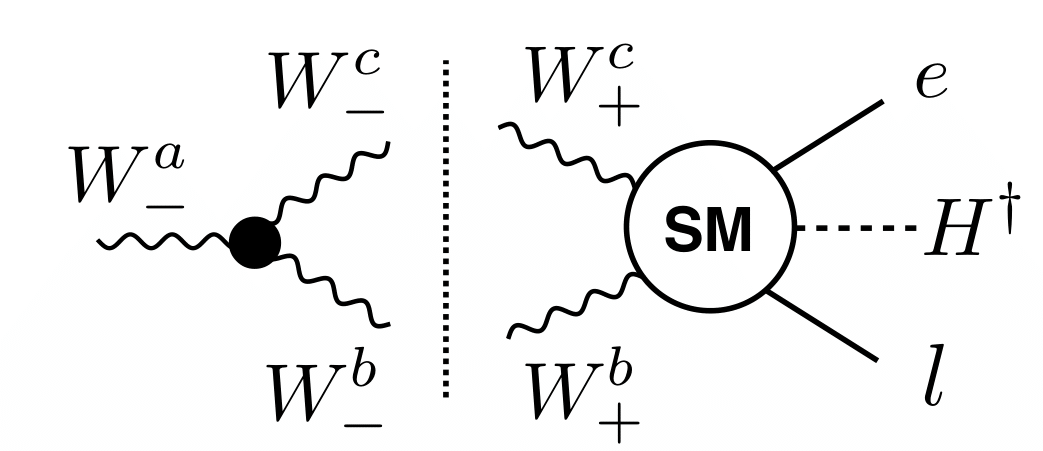}
\caption{\it Potential extra contribution from $C_{W^3}$  to the anomalous dimension of $C_{WHle}$.  This should be considered for  the anomalous dimension of the form-factor  ${\cal F}_{WHle}$, \eq{CH}  
 (where $p_b+p_c+p_a\not=0$), but not when using \eq{MF}.}
\label{W3todipole2}
\end{figure}

The only $n=3$ amplitude at order $1/\Lambda^2$ is  given in \eq{F3d}.  
In order to contribute to  $W^a_-H^\dagger le$, it must involve $W$ bosons:
\be
{\cal A}_{W^3}(1_{W^a_-},2_{W^b_-},3_{W^c_-})=\frac{iC_{W^3}}{\Lambda^2}\la 12\ra \la 23\ra \la 31\ra f^{abc}\,,
\label{W3d}
\ee
where $f^{abc}$ are the SU(2) structure constants.
Its    potential contributions to 
$W^a_-H^\dagger le $  are given by the two diagrams of  Fig.~\ref{W3todipole}. 
Although the contribution from Fig.~\ref{W3todipole2} should not be considered in \eq{MF} (it involves an $n=3$ amplitude), it would contribute if we were using \eq{CH}.  
We have calculated this contribution to ${\cal F}_{WHle}$ to check that, as expected,
it    smoothly  goes to zero as $p_a+p_b+p_c=Q\to0$, so that both \eq{CH} and \eq{MF} give the same result in this limit.

The LHS amplitudes of Fig.~\ref{W3todipole}  appear for the first time, and must be calculated. 
Interestingly,  they can be fully determined by just demanding proper factorization and
crossing $a\leftrightarrow b$. We obtain 
\bea
&(a)&  \widehat {\cal A}_{W^3}(3_{W^a_-},4_{H^\dagger},1'_{H},2'_{W^b_-})
=\frac{ig_2C_{W^3}f^{abc}T^c}{2\Lambda^2}\left[\frac{\la 31'\ra \la 42'\ra \la 32'\ra}{ \la 1'4 \ra}-\frac{\la 2'1'\ra \la 34\ra \la 32'\ra}{ \la 1'4 \ra}\right]\,,\ \ \ \ \ \ \label{F3toF2h2}\\
&(b)&  \widehat {\cal A}_{W^3}(3_{W^a_-},2_{{l}},1'_{\bar {l}},4'_{W^b_-})
=\frac{ig_2C_{W^3}f^{abc}T^c}{\Lambda^2}\frac{\la 34'\ra \la 32\ra \la 24'\ra}{ \la 1'2 \ra}\,.
\eea
With the above formulas and \eq{su2}, and after using a couple of times the Schouten identity \eq{Schouten} to 
reorder the indices  inside the brackets,
 we can write the RHS of  \eq{MF} as
\bea
&(a)& r\, T^a\int d{\rm LIPS}\,  \la 12\ra    \left[\frac{\la 32'\ra}{2}  \left(\frac{\la 31'\ra}{\la 2'1'\ra} + \frac{ \la 23\ra}{\la 2'2\ra}\right)
+\la 43\ra\left(\frac{\la 31'\ra}{\la 41'\ra}+\frac{\la 32\ra\la 1'2'\ra}{ \la 41'\ra \la 2'2\ra}\right)\right]
\,,\ \ \ \ \ \
\label{int1}\\
&(b)&  r\, T^a\int d{\rm LIPS}\,  \la 23\ra    \left[\la 34'\ra  \left(\frac{\la 11'\ra}{\la 4'1'\ra} + \frac{\la 41\ra}{\la 4'4\ra}\right)
+\la 21\ra\left(\frac{\la 31'\ra}{\la 21'\ra}+\frac{\la 34\ra\la 1'4'\ra}{\la 21'\ra \la 4'4\ra}\right)\right]\,.\ \ \ \ \ \ 
\label{int2}
\eea
where $ r=-\frac{g_2^2y_e C_{W^3}}{4\pi^3\Lambda^2}$, and we have used $f^{abc}T^bT^c={iN}T^a/2$ for SU(N) groups.
We now relate the internal primed spinors to the external ones. Specifically, to $| 1 \ra$ and $|2\ra $ in $(a)$, and to $| 1 \ra$ and $|4\ra $ in $(b)$. We use relations similar to \eq{34primed}, 
and get
\bea
&(a)&  \gamma_{WHle}
=-\frac{\pi r}{2} \left[\frac{1}{2} -\frac{\la 12\ra \la 34\ra }{\la 31\ra}\int^{\pi/2}_0d\theta\, 2{s_\theta}\int^{2\pi}_0 \frac{d\phi}{2\pi} \frac{1}{\la 42\ra c_\theta+\la 41\ra s_\theta e^{-i\phi}}\right], \ \ \ \ \ \ 
\label{W3todipolea}\\
&(b)&  \gamma_{WHle}= \frac{\pi r}{2}\left[\frac{1}{2}+\frac{\la 12\ra \la 34\ra }{\la 31\ra }\int^{\pi/2}_0d\theta\, 2{s_\theta}\int^{2\pi}_0 \frac{d\phi}{2\pi} \frac{1}{\la 42\ra c_\theta+\la 12\ra s_\theta e^{-i\phi}}\right].
\ \ \ \ \ \   \label{W3todipoleb}
\eea
The second term of Eqs.~(\ref{W3todipolea})--(\ref{W3todipoleb}) 
can be calculated using Cauchy's residue theorem:
\be
\int^{2\pi}_0 \frac{d\phi}{2\pi}\,  \frac{1}{a+e^{-i \phi}}=
\frac{1}{2\pi ia}\oint dz \frac{1}{z+1/a}=\frac{1}{a}\Theta\left(1-\left|\frac{1}{a}\right|\right)\,,
\ee
where the contour travels along the unit circle counterclockwise. This leads to  logarithmic terms, like
\be
(a)~~~\frac{\pi r}{2}\frac{\la 12\ra \la 34\ra }{\la 31\ra }\int^{\pi/2}_0\!\!d\theta\, 2{s_\theta}\int^{2\pi}_0 \frac{d\phi}{2\pi} \frac{1}{\la 42\ra c_\theta+\la 41\ra s_\theta e^{-i\phi}}=
\frac{\pi r}{2}
\frac{s_{12}}{s_{24}}\ln \frac{s_{14}}{s_{24}+s_{14}},
\label{logterm}
\ee
indicating the presence of box and triangle contributions (see Appendix~\ref{app1}).
Nevertheless, when adding   $(a)$ and $(b)$, the logarithms cancel out, as expected.
Surprisingly, also the constant terms, the first terms of Eqs.~(\ref{W3todipolea})--(\ref{W3todipoleb}),
cancel out, giving $\gamma_{WHle}=0$, as found previously in the literature 
\cite{Boudjema:1990dv}.

The above calculation of $\gamma_{WHle}$  can be  greatly simplified by using \eq{ACK} instead of \eq{MF}. The reason is that, as we explained, in \eq{ACK} triangle and box contributions are projected out with the $z$ integration.
Indeed, by performing the BCFW shifts $|1'\ra\to |1'\ra+ z |2'\ra$ and $|1'\ra\to |1'\ra+ z |4'\ra$ respectively in Eqs.~(\ref{int1})--(\ref{int2}),
and taking the constant term of the Laurent series at $z=\infty$, 
the last terms of Eqs.~(\ref{int1})--(\ref{int2})
go to zero, and only the constant terms of Eqs.~(\ref{W3todipolea})--(\ref{W3todipoleb}) remain.
This shows the usefulness of \eq{ACK}.

The above result can also be used for the contribution of a 3 Gluon ($G_-^a$) amplitude (similar 
to \eq{W3d}, but with $W\to G$)
to the chromodynamic down-quark dipole,  that is, to the amplitude $G^a_-H^\dagger qd$. 
In this case, only the diagram $(b)$ of Fig.~\ref{W3todipole} contributes, 
with $W\to G$, $l\to q$ and $e\to d$,
in addition to a similar  diagram obtained from the interchange $q\leftrightarrow d$.
The SM amplitude to use in this case is \eq{su3case}.
We find that, while the logarithmic terms cancel as expected, the constant term remains, giving
\be
\gamma_{GHqd}=\frac{3g_3^2y_d}{16\pi^2}C_{G^3}\,.
\label{GHqd}
\ee
\subsection{Comparison with the literature}

We can compare our results for the anomalous dimensions  with those reported in the literature,
mainly done using the Feynman diagrammatic approach
(see for example \cite{Elias-Miro:2013gya,Alonso:2013hga,Buchalla:2019wsc}).
For this purpose, we need to relate the dimension-6 operators of the SM EFT
to our amplitudes. This is presented in Appendix~\ref{app3}, in the basis of Ref.~\cite{Grzadkowski:2010es}.
Using these relations, we have checked that our calculations reproduce the anomalous dimensions
of the Wilson coefficient of the SU(2)$_L$ dipole operator ${\cal O}_{eW}=\bar L_L\sigma^a\sigma^{\mu\nu}e_R H W^a_{\mu\nu}$
found  in the literature (see for instance \cite{Alonso:2013hga}
and  \cite{Panico:2018hal})\footnote{For \eq{GHqd} we agree with \cite{Braaten:1990gq} and the errata of \cite{Alonso:2013hga}.}.
For the 4-fermion operators,
using the relations in Appendix~\ref{app3}
together with ${\cal O}_{lequ}^{(3)}= -8
(\bar L_{L\, i}  u_R) (\bar Q_{L\, j} e_R)\epsilon_{ij}-4{\cal O}_{lequ}^{(1)}$, where
${\cal O}_{lequ}^{(1)}=(\bar L_{L\, i}  e_R) (\bar Q_{L\, j} u_R)\epsilon_{ij}$,
we can relate the anomalous dimensions of \eq{glequ}
with those in Ref.~\cite{Alonso:2013hga}.
We find also agreement.
We would like again to emphasize 
the  similar  origin
of the   anomalous dimensions
of $C_{eW}$ and  $C_{lequ}^{(3)}$, 
made evident via the  on-shell method  discussed here,
that allowed for  non-trivial checks of contributions arising from
very different Feynman diagrams.

\section{Conclusion}
\label{sec:conclusion}

We  have  initiated here a systematic treatment of effective theories
via on-shell amplitudes, 
where the presence of new physics at some scale $\Lambda$ is encoded in  new
 ``elementary" amplitudes  ${\cal A}_{\mathcal{O}_i}$, suppressed by powers of $E/\Lambda$. 
This approach is an  alternative to the usual operator expansion  performed using Lagrangians.
Here, it is the coefficients $C_{\mathcal{O}_i}$ in front of the amplitudes that play the role of the Wilson coefficients.

The on-shell approach has  several advantages. For instance, it avoids the usual problems with redundancies present   in the Lagrangian approach, and also makes it  much easier to understand 
the   physical implications of the theory.
Furthermore,   it allows the use of generalized unitarity methods to obtain information about the 
quantum structure of the theory, without the need of explicitly performing  one-loop calculations.

The main purpose of this article has been  to show the effectiveness of on-shell techniques in
computing the   anomalous dimensions of  $C_{\mathcal{O}_i}$.
We have done  this by considering many examples in the SM  at order $E^2/\Lambda^2$.\footnote{Of course, the use of these techniques is not limited to the SM. The same authors have used them for example to investigate some properties of the chiral theory for pions
at the one-loop order \cite{talks}.} 
In particular, we have calculated all anomalous dimensions (except for the self-renormalization) of  the dipole coefficient $C_{WHle}$ defined in \eq{treelevel}.
We have shown how one can  calculate anomalous dimensions from \eq{MF}, 
that corresponds to  just   sewing together two tree-level on-shell amplitudes via an integration over a  two-particle phase-space. 
This integral   can be reduced to an angular integration that in most cases reveals to be trivial. 
Apart from the unavoidable intricacies coming from the fact that there are many different species of particles in the SM, the on-shell method shows a remarkable simplicity.
In particular, several simple selection rules \cite{Cheung:2015aba,Bern:2019wie} such as \eq{sr}, 
but also new ones derived in this paper (see \eq{ADpsi4}),
help to understand certain non-renormalizations.

Moreover, we have seen  that the method is quite efficient,  
as it  requires  the calculation of  only a few  SM   amplitudes,
from which one  can  deduce many different anomalous dimensions.
This   recycling advantage  
has allowed  to relate    $\gamma_i$'s that in the Feynman approach originate from very different diagrams. In particular, the renormalization of $C_{WHle}$  from $C_{lequ}$ and 
$C_{luqe}$  can be related to its inverse: 
the renormalization of $C_{lequ}$ and 
$C_{luqe}$ from $C_{WHle}$.
This has provided    non-trivial checks of previous results in the literature.

In some cases ($\Delta n\geq 1$), we have seen that the phase-space integral is less trivial and leads to logarithms of ratios of Mandelstam invariants.
Nevertheless, these logarithmic terms, which appear in the individual contributions to $\gamma_i$ but cancel in the total sum, can be easily avoided through a refined
sewing procedure, \eq{ACK}, that includes a simple contour integral (which essentially amounts to performing a trivial Taylor expansion around complex infinity).
In Appendix \ref{app1}, we have explored what is behind the emergence of these logs. We show that they are due to the presence of box topologies in the loop amplitude.
We have also found that the  cancellation of the logarithms in the anomalous dimensions is guaranteed by the absence of IR divergencies in the process.

Here, we have worked under a couple of assumptions: $(i)$ that no IR divergencies are involved 
and $(ii)$ that in the renormalization of an amplitude,  
only one type of $1/\Lambda^2$ amplitude  $\mathcal{A}_{\mathcal{O}_i}$  appears at tree-level.
We hope to report soon on the more general situation.

\vspace{1.0cm}

{\it Note added:} After submitting  this manuscript we became aware of the work of \cite{EliasMiro:2020tdv}  in which  the on-shell method  is also  considered 
to calculate anomalous dimensions in the SM EFT. See also the companion article
\cite{Baratella:2020dvw}.

\section*{Acknowledgments}
We would like to thank Benedict von Harling for discussions. We also thank Aneesh Manohar and
Rodrigo Alonso for correspondence concerning the SM EFT anomalous dimensions.
P.B. has been partially supported by the DFG Cluster of Excellence 2094 ORIGINS, the Collaborative Research Center SFB1258, the BMBF grant 05H18WOCA1, and thanks  for the hospitality the Munich Institute for Astro- and Particle Physics (MIAPP), which is funded by the Deutsche Forschungsgemeinschaft (DFG, German Research Foundation) under Germany’s Excellence 
Strategy - EXC-2094 - 390783311. C.F. is supported  by the fellowship FPU18/04733 from the Spanish Ministry of Science, Innovation and Universities. A.P. is supported by the Catalan ICREA Academia Program and  grants FPA2017-88915-P, 2017-SGR-1069 and SEV-2016-0588.

\appendix
\section{Cancellation of IR divergencies and absence of triangle and box contributions in the sum over 2-cuts}
\label{app1}

\begin{figure}[t]
\centering
\includegraphics[width=.3\textwidth]{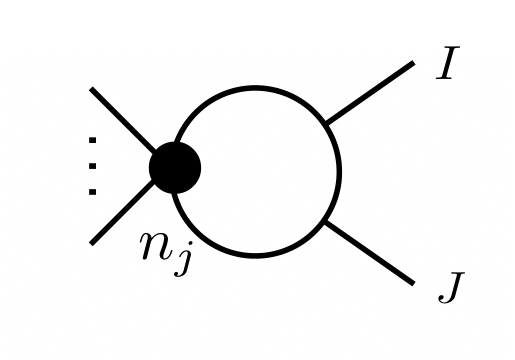}
\caption{\it  One-loop contribution from an amplitude with $n_j$ legs to  the renormalization of an amplitude   with $n_i=n_j$ legs.}
\label{deltan0}
\end{figure}

In this Appendix, we consider 
one-loop mixings ${\cal A}_{\mathcal{O}_j}\rightarrow {\cal A}_{\mathcal{O}_i}$ 
having $\Delta n\equiv n_i-n_j=0,1$. By exploiting the properties of \eq{general}, we show that
triangles and boxes do not  contribute to \eq{MF}.
The proof relies in the absence of IR divergencies, so it is only valid when $\gamma_{\rm IR}=0$.

Let us start by considering the case $\Delta n=0$. Apart from bubble integrals, which are IR safe,
we can also have triangles, as shown
in Fig.~\ref{deltan0}. The reason why boxes are absent is topological:
there are simply not enough external legs to make them.
The relevant triangle integrals are of the following form \cite{Britto:2010xq}:
\be
I_3^{(IJ)}=\frac{\alpha(\epsilon) \mu^{2\epsilon}}{\epsilon^2}(-s_{IJ})^{-1-\epsilon}\,,\\
\label{I3}
\ee
where $s_{IJ}=(p_I+p_J)^2$ and 
 \be
 \alpha(\epsilon)= \frac{\Gamma(1+\epsilon) \Gamma^2(1-\epsilon)}{\Gamma (1 -2\epsilon){ (4\pi)^{\frac{D}{2}}}}=\frac{1}{16\pi^2}+O(\epsilon)\,.
 \ee
We use here $I,J,\ldots$ indices for particle labels to help avoiding confusion. In dimensional regularization, the $\epsilon^{-2}$ pole in \eq{I3} signals that the integral is IR divergent. In fact, on dimensional grounds, we know that it is convergent in the UV.  
Expanding for $\epsilon\to 0$, we have
\be
\alpha(\epsilon)^{-1}\,
I_3^{(IJ)}\to-\frac{1}{s_{IJ}} \left(\frac{1}{\epsilon^2}-\frac{1}{\epsilon}\ln\left(\frac{-s_{IJ}}{\mu^2}\right)
\right)+O(1)\,.
\label{1mI3}
\ee
Since the IR divergence of the full amplitude {is zero by assumption}, we have 
the following conditions:
\be
\sum_{I,J} \frac{C_3^{(IJ)}}{s_{IJ}}=0\ ,\ \ \ \ \ \
\sum_{I,J} \frac{C_3^{(IJ)}}{s_{IJ}}\ln(-s_{IJ})=0\,,
\label{conditions}
\ee
where we sum over all the distinct triangle topologies. The two conditions come, respectively, from the
cancellation of the $\epsilon^{-2}$ and $\epsilon^{-1}$ poles.
Even though the first condition could be satisfied for a nontrivial configuration of the triangle coefficients,
the second one requires $C_3^{(IJ)}=0$ for all $I,J$. The reason is that the logarithms $\ln(-s_{IJ})$
cannot be canceling among themselves, unless trivially some of the $s_{IJ}$ are equal (see below for the case $n_i=4$).
Technically, this is because the $C_3$'s are rational functions of the kinematical variables,
while the logarithms are transcendental.

The cases $n_i=3,4$ are special.
For three particles, $s_{IJ}=0$ for each $I,J$, implying that all triangle integrals $I_3^{(IJ)}$ are scaleless and vanish.
On the other hand, in the 4-particle case we have 
 $s_{12}=s_{34}$,  $s_{13}=s_{24}$ and  $s_{14}=s_{23}$,
 and we cannot exclude the nontrivial configuration $C_3^{(12)}=-C_3^{(34)}$, $C_3^{(13)}=-C_3^{(24)}$ and $C_3^{(14)}=-C_3^{(23)}$. 
Nevertheless, all the triangle contributions, including the finite parts, cancel in pairs. For example
\be  
C_3^{(12)} I^{(12)}_3+C_3^{(34)}I^{(34)}_3=0\,.
\ee
Either way, we see that the total triangle contribution is required to be zero in order to have an IR-safe amplitude.
This means in particular that no triangle (nor box) contribution can appear in \eq{MF}.

\begin{figure}[t]
\centering
\includegraphics[width=.7\textwidth]{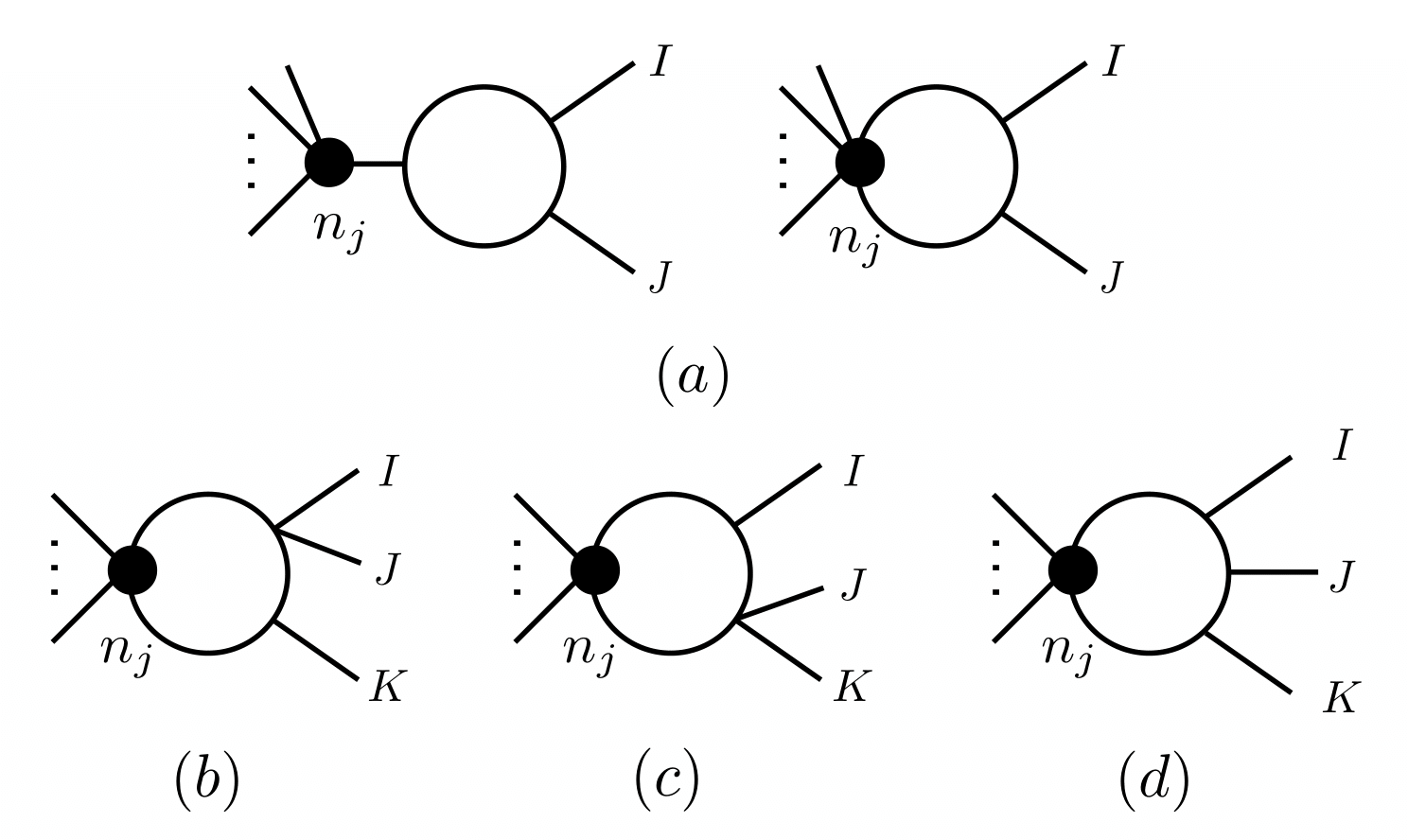}
\caption{\it  One-loop contributions from an amplitude with $n_j$ legs to  the renormalization of an amplitude   with $n_i=n_j+1$ legs.}
\label{deltan1}
\end{figure}

We now move to the case  $\Delta n=1$.
With one additional external leg, we can build new one-loop topologies,
as displayed in Fig.~\ref{deltan1} (we show only those topologies which are associated to IR divergent integrals).
We have triangles, like  $(a)$ and $(b$-$c)$, and boxes as well $(d)$.
The corresponding integrals \cite{Britto:2010xq} are given, respectively, by \eq{1mI3} and 
\bea
I_3^{(IJ|K)}&=&\frac{ \alpha(\epsilon) \mu^{2\epsilon}}{\epsilon^2}\frac{(-s_{IJ})^{-\epsilon}-(-s_{IJK})^{-\epsilon}}{(-s_{IJ})-(-s_{IJK})}\ , \ \ \ \ \ \ I_3^{(I|JK)}=I_3^{(IJ|K)}(I\leftrightarrow K)\,,\\
I_4^{(IJK)}&=&\frac{ \alpha(\epsilon) \mu^{2\epsilon}}{\epsilon^2}\frac{2}{s_{IJ}s_{JK}} \left[(-s_{IJ})^{-\epsilon}+(-s_{JK})^{-\epsilon}-(-s_{IJK})^{-\epsilon}\right]-\frac{1}{16\pi^2}F_4^{(IJK)}\,,   \label{epsbox}
\eea
where $s_{IJK}=(p_I+p_J+p_K)^2$ and
\be
F_4^{(IJK)}=\frac{2}{s_{IJ}s_{JK}}\left[{\rm Li}_2\left( 1-\frac{s_{IJK}}{s_{IJ}} \right)+{\rm Li}_2\left( 1-\frac{s_{IJK}}{s_{JK}} \right)+\frac{1}{2}\ln^2\left(\frac{s_{IJ}}{s_{JK}}\right)+\frac{\pi^2}{6}\right]+O(\epsilon)\,.
\label{noepsbox}
\ee
We refer to Fig.~\ref{deltan1} for the notation.

In this case, the cancellation of IR divergencies could be nontrivial, occurring between triangles and boxes and thus implying
a relation among their coefficients, $C_3$ and $C_4$. 
In other words, triangles and boxes could appear in combinations 
free from IR divergencies, as for example
\be
 s_{IJ}s_{JK}I_4^{(IJK)}+  s_{IJ} I_3^{(IJ)}+ s_{JK} I_3^{(JK)}+(s_{IJ}-s_{IJK})I_3^{(IJ|K)}+(s_{JK}-s_{IJK})I_3^{(I|JK)},
\label{IRsafe}
\ee
which is proportional to $s_{IJ}s_{JK}F_4^{(IJK)}$. 
 Imposing this condition,    $\mathcal{A}_{\rm loop}$ reduces to
\be
\mathcal{A}_{\rm loop}=\sum_a C_2^{(a)} I_2^{(a)}-\frac{1}{16\pi^2}\sum_c C_4^{(c)}F_4^{(c)},
\label{loopnoIR}
\ee
that shows that a finite contribution  from boxes remains in the one-loop amplitude. 
Nevertheless, as we now prove, this second term of \eq{loopnoIR}  
does not contribute to the sum over 2-cuts.

A 2-cut of an amplitude is computed with the Cutkosky rule,  that consists in substituting the loop propagators $\ell^{-2}$  and $(\ell-P)^{-2}$ with  respectively $\delta^+(\ell^2)$ and $\delta^+(\ell^2-P)$.
We normalize the 2-cuts in such a way that the 2-cut of a bubble gives
\be
{\rm Cut}_2[I_2^{(a)}]=-\frac{1}{8\pi^2}\,.
\ee
Summing over all possible 2-cuts of  \eq{loopnoIR}, 
and using  \eq{gamma} with $\gamma_{\rm IR}=0$,
we obtain
\be
\sum_{\rm 2-cuts} {\rm Cut}_2[\mathcal{A}_{\rm loop}]=\gamma_i \mathcal{A}_{{\cal O}_i} -\frac{1}{16\pi^2}\sum_{c} C_4^{(c)}\sum_{\rm 2-cuts}{\rm Cut}_2[F_4^{(c)}]\,.
\label{totalsum}
\ee
We want to prove that the second term in \eq{totalsum} vanishes.
For the box contribution $(d)$ of Fig.~\ref{deltan1}, we have 
three possible nonzero 2-cuts, 
corresponding to cutting out either $(IJ)$, $(JK)$ or  $(IJK)$ from the rest of  states.
By applying the Cutkosky rule to the IR-safe combination in \eq{IRsafe}, we can deduce that
these three 2-cuts give, respectively,
\bea
{\rm Cut}_2^{(IJ)}[F_4^{(IJK)}]&=&\frac{4}{ s_{IJ}s_{JK}}\ln\left(\frac{s_{IJK}-s_{IJ}}{s_{JK}}\right)\,, \label{cut12}\\
{\rm Cut}_2^{(JK)}[F_4^{(IJK)}]&=&\frac{4}{ s_{IJ}s_{JK}}\ln\left(\frac{s_{IJK}-s_{JK}}{s_{IJ}}\right)\,,
\label{cut23}\\
{\rm Cut}_2^{(IJK)}[F_4^{(IJK)}]&=&\frac{4}{ s_{IJ}s_{JK}}\ln\left(\frac{s_{IJ}s_{JK}}{(s_{IJK}-s_{JK})(s_{IJK}-s_{IJ})}\right)\,.
\label{cut123}
\eea
Crucially, these three terms add up to  zero.
This completes the proof that, for IR-finite processes with $\Delta n=1$,
triangle and box contributions vanish in the total sum over 2-cuts.

We stress that  box contributions to   individual 2-cuts do not have to be zero,
and therefore  logarithms (those of Eqs.~(\ref{cut12})--(\ref{cut123})) can be present  before the total sum
is performed.
An example of this phenomenon is \eq{logterm}. 
Next, we check that it is precisely $C_4$ that fixes the coefficient of the logarithm
in \eq{logterm}.

As a final comment, we observe that, for $n_i=4$, the third cut \eq{cut123} vanishes, since $s_{IJK}=0$.
In fact, in this case the 2-cut is massless.

\subsection{Box contributions from quadruple cuts}

\begin{figure}[t]
	\centering
	\includegraphics[width=0.5\textwidth]{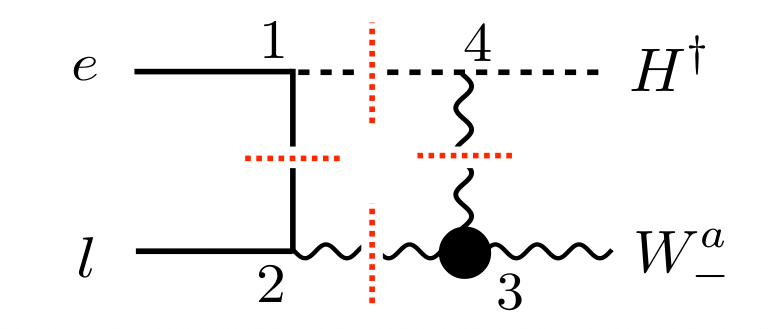}
	\caption{\it Quadruple cut in ${\cal A}_{W^3}\to{\cal A}_{WHle}$.}
	\label{4cut}
\end{figure}

Here, we calculate the (unique) box contribution to the one-loop renormalization ${\cal A}_{W^3}\to{\cal A}_{WHle}$,
discussed in Section~\ref{W3}. Due to the presence of logarithms in \eq{logterm}, and according to the results presented just before,
we expect indeed a  nonzero coefficient $C_4$. 

We will follow Ref.~\cite{ArkaniHamed:2008gz},
where the box contribution is calculated from a  quadruple cut (4-cut).
Since the relevant amplitude has four external states,
after a 4-cut it reduces to a product of four $n=3$  amplitudes
(see Fig.~\ref{4cut}), which are completely fixed by the little group. 
We have
\be
C_4=\frac{1}{2}{\cal A}_1(p_1,\ell^{+}_{41},-\ell^{-}_{12}) {\cal A}_2(p_2,\ell^{-}_{12},-\ell^{+}_{23}) {\cal A}_3(p_3,\ell^{+}_{23},-\ell^{-}_{34}) {\cal A}_4(p_4,\ell^{-}_{34},-\ell^{+}_{41})+(-\leftrightarrow +)\,,
\label{c4}
\ee
where   $\ell^{\pm}_{ij}$ defines the momentum that goes from vertex $i$ to vertex $j$
 and, as explained in \cite{ArkaniHamed:2008gz}, we have
two possible sets, labelled by the $\pm$.
The two $\ell^{\pm}_{ij}$  are related by complex conjugation, that is $\ell_{ij}^-=(\ell_{ij}^+)^*$,
and can be elegantly written in terms of spinor-helicity variables, as found in Ref.~\cite{Johansson:2012sf}. For example, we have
\be
\ell^+_{12}=\frac{\langle 23\rangle}{\langle 31\rangle} | 2 ]\langle 1 |\ ,\ \ \ \ \ 
\ell^-_{12}=\frac{[ 23]}{[ 31]} | 1]\langle 2 |\,,
\label{linabc}
\ee
with similar expressions for the other cut  momenta $\ell^\pm_{ij}$, obtained by cyclic permutation of the labels.
One can check that they satisfy the on-shell condition $(\ell^\pm_{ij})^2=0$, and relations like $\ell_{12}^+ +p_2=\ell_{23}^-$ (which explain the choices in \eq{c4}).

Let us now move to compute
the 4-cut in the  process ${\cal A}_{W^3}\to{\cal A}_{WHle}$, as represented in Fig.~\ref{4cut}.
The relevant $n=3$ amplitudes are given in \eq{W3d} and Eqs.~(\ref{3pti})--(\ref{3ptf}):
\be
{\cal A}_1={ i}y_e\langle 1 \ell_{12}\rangle\ ,\ \  \ \ 
{\cal A}_2=g_2\frac{[\ell_{12}\ell_{23}]^2}{[\ell_{12}2]}(T^b)_{kj}\,,
\ee
\be
{\cal A}_3=\frac{i C_{W^3}}{\Lambda^2} \langle\ell_{23}3\rangle \langle 3\ell_{34}\rangle \langle \ell_{34}\ell_{23}\rangle f^{abc}\ ,\ \ \ \
{\cal A}_4=g_2\frac{[4\ell_{34}][\ell_{41}\ell_{34}]}{[\ell_{41}4]}(T^c)_{ik}\,.
\ee
The product ${\cal A}_1\ldots {\cal A}_4$
 can be manipulated in order to reduce the number of 
$\ell_{ij}$ spinors. We find 
\be
{\cal A}_1 {\cal A}_2 {\cal A}_3 {\cal A}_4={ i}\frac{g_2^2y_eC_{W^3}}{\Lambda^2}s_{12}\langle 23\rangle \langle 12 \rangle [2 |\ell_{23}| 3 \rangle(T^a)_{ij}\,.
\label{pro}
\ee
Then, by making use of \eq{linabc} we get (notice that only $\ell_{23}^+$ contributes to \eq{pro})
\be
C_4=-\frac{g_2^2y_eC_{W^3}}{2\Lambda^2}(T^a)_{ij} \langle 31\rangle \langle 32 \rangle \frac{s_{12}^2 s_{23}}{s_{13}} =-\frac{g_2^2y_eC_{W^3}}{2}
\frac{s^2_{12}s_{23}}{s_{13}}{\cal A}_{WHle}\,,
\label{C4final}
\ee
{where we also multiplied by $-i$ due to the internal fermion line in Fig.~\ref{4cut}, as explained in \ref{app2a}}.
We now want to use the above result to obtain \eq{logterm},
which corresponds to taking a 2-cut in the $(12)$-channel (see Fig.~\ref{W3todipole}, $(a)$).
Using \eq{cut12} with $I=1,J=2$ and $K=3$, we get
\be
{\rm Cut}^{(12)}[\mathcal{A}_{\rm loop}]=-\frac{C_2^{(12)}}{8\pi^2}-\frac{C_4}{4\pi^2 s_{12}s_{23}}\ln\left(\frac{-s_{12}}{s_{23}}\right)\subseteq \gamma_{WHle} \mathcal{A}_{WHle}\,.
\label{logfromcut}
\ee 
After dividing  by $\mathcal{A}_{WHle}$ and using \eq{C4final},  we find that \eq{logfromcut} 
agrees with \eq{logterm}.

\section{SM on-shell amplitudes}
\label{app2}

\subsection{Conventions}
\label{app2a}

We start with the conventions taken in this article.
We choose the metric $\eta_{\mu\nu}={\rm diag}(+,-,-,-)$,
and the 2-component spinors with $h=\mp 1/2$ to be denoted respectively by $|p\ra _\alpha$ and $|p]^{\dot\alpha}$. The momentum is given by $p_{\alpha\dot\alpha}=|p\ra _\alpha [p|_{\dot\alpha}$, and the contractions are
\be
\la pq\ra \equiv \la p|^\alpha |q \ra_\alpha\ \  \ \  {\rm and}\ \  \ \   
[ pq] \equiv [p|_{\dot\alpha} |q]^{\dot\alpha}\,,
\label{contractions}
\ee
where we follow the conventions of Ref.~\cite{Dreiner:2008tw}  for raising and lowering indices.
We also define $\la i|\sigma_\mu|j]\equiv \la i|^\alpha  (\sigma_\mu)_{\alpha\dot\alpha}|j]^{\dot\alpha}$, 
that fulfill  the property $\la i|\sigma_\mu| j]=[j|\sigma_\mu|i\ra$.
We also have
\be
p_i^\mu=\frac{1}{2}\la i|\sigma^\mu|i]\ , \ \ \ \ \ 
2\, p_i\cdot p_j=\la ij\ra [ji]\,,
\ee
the  Fierz relation
\be
\la i|\sigma^\mu| j]\la k|\sigma^\mu|l]=-2\la ik\ra [jl]\,,
\ee
and the Schouten identity
\be
\la ij\ra \la kl\ra=\la ik\ra \la jl\ra-\la il\ra \la jk\ra\,.
\label{Schouten}
\ee
We are considering amplitudes with all states incomming. Outgoing states can be related to incomming states with opposite momentum and helicity, replacing particle $\leftrightarrow$ antiparticle. 
When we encounter spinors with negative momenta,  it is convenient to write them back with positive momenta. Following  the appendix of Ref.~\cite{Mangano:1990by}, we define
\be
|-p\ra _\alpha=i|p\ra _\alpha\, \ , \ \ \ |-p]^{\dot\alpha}=i|p]^{\dot\alpha}\,,
\label{signs}
\ee
that  consistently leads to $|-p\ra [-p|=-p$.

\begin{figure}[t]
	\centering
	\includegraphics[width=0.4\textwidth]{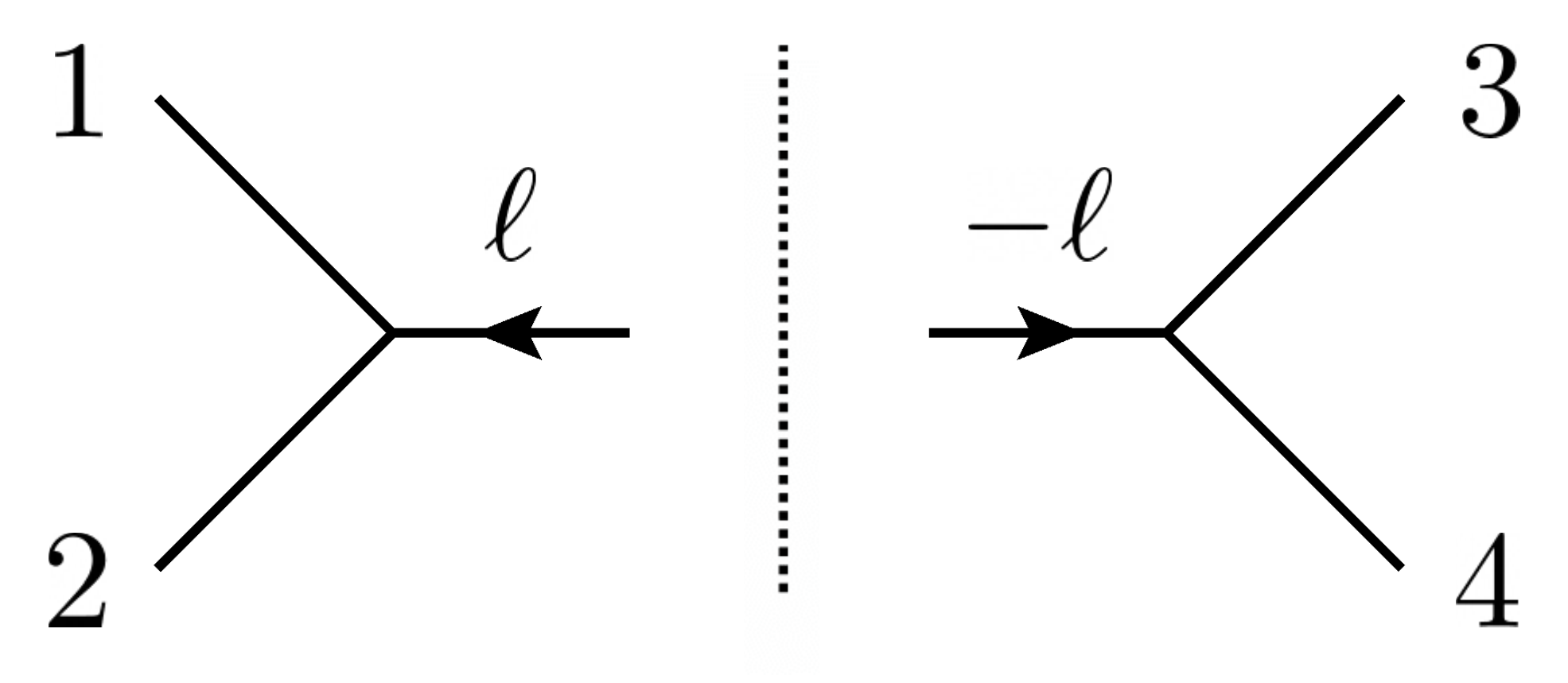}
	\caption{\it 
		Diagram of the factorization of ${\cal A}(1,2,3,4)$ into the product of 3-point subamplitudes.}
	\label{ifact2}
\end{figure}

The convention \eq{signs} fixes the factorization of an amplitude ${\cal A}$ into a product of lower-point amplitudes. For the case of a 4-point amplitude with a pole in the $s$-channel (see Fig.~\ref{ifact2}), this is given by
\be
\lim_{s_{12}\rightarrow 0}s_{12}{\cal A}(1,2,3,4)=i^{F[\ell]} \ i{\cal A}_L(1,2,\ell)\ i{\cal A}_R(-\ell,3,4)\, ,
\label{factorization}
\ee
where $\ell=p_3+p_4$ and $F[i_1...i_n]$ counts the number of fermions or antifermions in the list $\{i_1,...,i_n\}$.
In case the amplitude has a pole in another channel, one has to reorder the particles, which may lead to an additional minus sign if there is an odd number of fermion exchanges. For example, if the pole is in the $t$-channel, we have
\be
\lim_{s_{13}\rightarrow 0}s_{13}{\cal A}(1,2,3,4)=i^{F[\ell]}(-1)^{n_{23}}\, i{\cal A}_L(1,3,\ell)\  i{\cal A}_R(-\ell,2,4)\,,
\ee
where $n_{23}=1$ if both particles 2 and 3 are fermions or antifermions, and 0 otherwise.

The need of the $i^{F[\ell]}$ factor in \eq{factorization}  can be understood by looking at the internal propagator. Amplitudes with fermions contain the following spinors 
\be
u_\mp(p)=
P_\mp\left(\begin{array}{c}|p\ra_\alpha \\|p]^{\dot\alpha} \end{array}\right)
\ ,\ \  \ \ 
\bar v_\mp(p)=
\big(\la p|^\alpha\  [p|_{\dot\alpha}\big)P_\mp\,,
\label{uv}
\ee
respectively for incoming $h=\mp1/2$ fermions and antifermions, where $P_\mp=(1\pm\gamma_5)/2$.
When factorizing an amplitude with an internal fermion,  there is a factor $u_\mp(\ell)$ in one of the subamplitudes and a factor $\bar{v}_\pm(-\ell)$ in the other one. Summing over all possible helicities, we have
\be
u_+(\ell)\bar v_-(-\ell)+u_-(\ell)\bar v_+(-\ell)=
i\sum_h u_h(\ell)\bar u_h(\ell)=
i \Slash \ell\,,
\ee
that leads to an  extra $i$ from the expected $\Slash \ell$. This is compensated with the $i$ factor in \eq{factorization} ({up to a minus sign that has to do with the fermion ordering in \eq{factorization}}). For vectors, however, the situation is different. The polarizations for incoming vectors with momentum $p$ are given by
\be
\epsilon_\mu^+=\frac{\la q|\sigma_\mu|p\ra}{\sqrt{2}\la qp\ra}
\ ,\ \  \ \ 
\epsilon_\mu^-=-\frac{\la p|\sigma_\mu|q]}{\sqrt{2} [qp]}\,,
\label{eps}
\ee
where $q$ is a  reference momentum  \cite{Dixon:2013uaa}. 
When considering an internal vector in \eq{factorization}, the polarizations come with  opposite sign for the momentum
in each subamplitude ${\cal A}_L$ and ${\cal A}_R$. Therefore we have
\be
\epsilon_\mu^+(\ell)\epsilon_\nu^-(-\ell)+\epsilon_\mu^-(\ell)\epsilon_\nu^+(-\ell)
=\sum_h \epsilon^h_\mu(\ell)(\epsilon^h_\nu(\ell))^*\,,
\label{recons}
\ee
where we have used  \eq{signs} and \eq{eps}.
\eq{recons}  gives  the proper sum over vector polarizations that we expect in a propagator, without having to add extra factors.

It is important to notice that the sign of the internal momenta $\ell$ is fixed as shown in Fig.~\ref{ifact2}. If we take the opposite momentum, the factor $i^{F[\ell]}$ has to be replaced by $\left(-i\right)^{F[\ell]}$. We can see this with an explicit example. Let us consider the amplitude ${\cal A}(1_e,2_{H^\dagger},3_{\bar{e}},4_H)$ in a Yukawa theory, ${\cal L}_y=-y_e H^\dagger el+h.c.$, where $e$ and $l$    are Weyl spinors of $h=-1/2$.
 Using \eq{factorization}, the 4-point amplitude can be computed as a product of two 3-point amplitudes, namely
\be
{\cal A}(1_e,2_{H^\dagger},3_{\bar{e}},4_H)=\frac{i}{s_{12}}\, i{\cal A}_L(1_e,2_{H^\dagger},\ell)\ i{\cal A}_R(-\ell,3_{\bar{e}},4_H)\,,
\label{fact0}
\ee	
that leads to
\be
{\cal A}(1_e,2_{H^\dagger},3_{\bar{e}},4_H)=-\frac{iy_e^2\la 1\ell\ra [-\ell 3]}{s_{12}}
	=-\frac{i^2y_e^2\la 1\ell\ra [\ell3]}{s_{12}}=y_e^2\frac{\la 14\ra}{\la 34\ra}\,.
\label{fact2}
\ee
Nevertheless, if we choose the momentum $\ell$ with an opposite sign, $\ell=-\left(p_3+p_4\right)$, 
we  have to use $\left(-i\right)^{F[\ell]}$ to obtain the same result:
\be
\frac{(-i)}{s_{12}}\, i{\cal A}_L(1_e,2_{H^\dagger},-\ell)\ i{\cal A}_R(\ell,3_{\bar{e}},4_H)=\frac{iy_e^2\la 1\!-\!\ell\ra [\ell3]}{s_{12}}
	=\frac{i^2y_e^2\la 1\ell \ra [\ell 3]}{s_{12}}=y_e^2\frac{\la 14\ra}{\la 34\ra}\,.
\label{fact1}
\ee

\begin{figure}[t]
	\centering
	\includegraphics[width=0.7\textwidth]{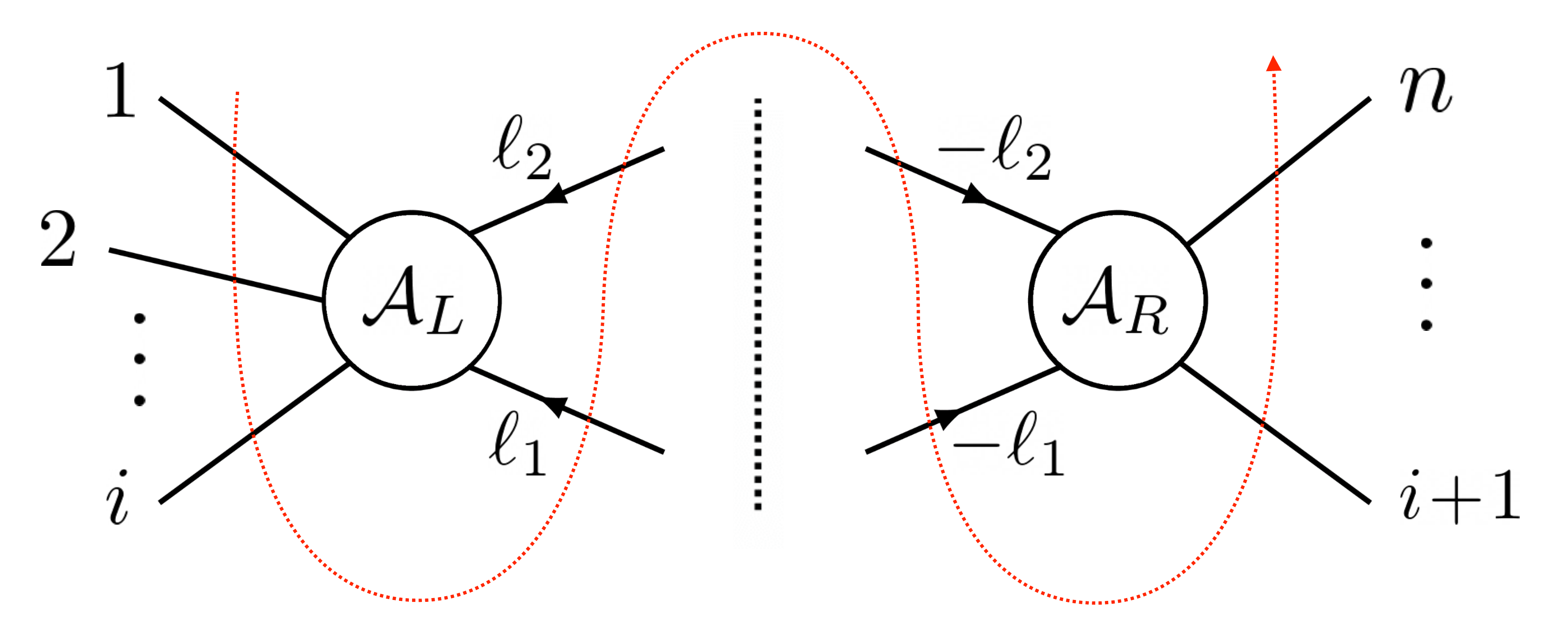}
	\caption{\it 
		Diagram of the double cut of a loop amplitude ${\cal A}(1,2,...,i,i+1,...,n)$. The red dotted line indicates the order in which one has to put the  fermions in the amplitudes.}
	\label{ifactloop}
\end{figure}

The generalization of \eq{factorization} to loop amplitudes  is straightforward.
When the internal lines go on-shell,  we need to add a factor $i$ per each fermion line.
For example, let us  consider the 2-cut of the loop amplitude of Fig.~\ref{ifactloop}. We  have
\be
{\cal A}(1,2,..., i,i+1,...,n)\rightarrow 
\int d{\rm LIPS}\ i^{F[\ell_1,\ell_2]} {\cal A}_L(1,2,...,i,\ell_1,\ell_2)\ {\cal A}_R(-\ell_2,-\ell_1,i+1,...,n)\,,
\ee
that leads to  \eq{MF}. Notice that the order of the states is important for fermions, that must follow the red dotted line as in
Fig.~\ref{ifactloop}.


\subsection{SM Amplitudes}

The on-shell amplitude approach  is based on building higher-point amplitudes from already existing ones of lower $n$.
The basic ``blocks'' are the  $n=3$ amplitudes, which are totally fixed by their helicities. For the SM
gauge boson interactions,
using the indices  $a,b,...$  for the adjoint  representation of the non-abelian groups,
and  $i,j$ indices 
for the fundamental representation,  we have
\bea
{\cal A}_{\rm SM}(1_{\psi_j}, 2_{\bar \psi_i},3_{V^a_-})=g_a\frac{\la 13\ra^2}{\la 12\ra} (T^a)_{ij}
\ ,\ \ \ \ \ \ \ \ \ \ \ \
{\cal A}_{\rm SM}(1_{\psi_j}, 2_{\bar \psi_i},3_{V^a_+})=g_a\frac{[23]^2}{[12]} (T^a)_{ij}\,,
\ \ \ \,&&
\label{3pti}\\
{\cal A}_{\rm SM}(1_{H_j}, 2_{H^\dagger_i},3_{V^a_-})=g_a\frac{\la 13\ra \la 23\ra}{\la  21\ra}(T^a)_{ij}
\ ,\ \ \ \
{\cal A}_{\rm SM}(1_{H_j}, 2_{H^\dagger_i},3_{V^a_+})=g_a\frac{[13][23]}{[12]}(T^a)_{ij}\,.&&
\label{3ptint}
\eea
For the abelian U(1)$_Y$ hypercharge we have  similar  expressions, with 
 $(T^a)_{ij}\to Y_i\delta_{ij}$.
We  fix our  normalization as ${\rm Tr}[T^a T^b]=\delta^{ab}/2$, with $Y_H=1/2$
and real $g_a$.
Let us comment that, in fact, only one of the amplitudes in Eqs.~(\ref{3pti})--(\ref{3ptint}) is enough to fix the definition of   the SM gauge coupling.
For instance, once the gauge interaction to a fundamental fermion is defined, the gauge interaction to scalars can be determined by the   consistency condition that $n>3$ amplitudes  must factorize into products of $n=3$ amplitudes (this is the equivalent to gauge invariance in the Lagrangian approach -- see for example \cite{Arkani-Hamed:2017jhn}).
Also, the second amplitudes in Eqs.~(\ref{3pti})--(\ref{3ptint})  can be determined from the first using CPT invariance and  unitarity.\footnote{Unitarity  $S^\dagger S=1$, where $S=1+iT$
and $T$ can be treated as a small perturbation around the identity,
 is needed to derive $T= T^\dagger+O(T^2)$ and therefore
${\cal A}=\la \alpha|T|\beta\ra\simeq\la \beta|T|\alpha\ra^*$.}

We also have  Yukawa interactions, that  for one family are given by (showing only  the SU(2)$_L$ indices)
\be
{\cal A}_{\rm SM}({1_{e}, 2_{l_i}},3_{H_i^\dagger})=y_e  \la 12\ra\ ,\ \ \  
{\cal A}_{\rm SM}({1_{d}, 2_{q_i}},3_{H_i^\dagger})=y_d  \la 12\ra\ ,\ \ \  
{\cal A}_{\rm SM}({1_{u}, 2_{q_i}},3_{H_j})=y_u  \la 12\ra\epsilon_{ij}\,.
\label{3ptfa}
\ee
These amplitudes fix our definitions of the SM Yukawa couplings $y_\psi$, that for one family 
can be taken to be real.
The generalization to 3 families is straightforward.
By CPT invariance and unitarity, we obtain
\be
{\cal A}_{\rm SM}({1_{\bar e}, 2_{\bar l_i}},3_{H_i})=y_e  [12]\ ,\ \ \ \ 
{\cal A}_{\rm SM}({1_{\bar d}, 2_{\bar q_i}},3_{H_i})=y_d  [12]\ ,\ \ \ \ 
{\cal A}_{\rm SM}({1_{\bar u}, 2_{\bar q_i}},3_{H_j^\dagger})=y_u  [ 12]\epsilon_{ij}\,.
\label{3ptf}
\ee
The relation between our gauge and Yukawa couplings, defined via amplitudes, and 
 the usual definitions arising from Lagrangians is provided in Appendix \ref{app3}.

From the above  $n=3$ amplitudes, we can build  $n=4$ amplitudes. Here, we quote the ones that are needed for this work.
These are   $V_+H\psi\psi$ amplitudes:
\be
{\cal A}_{\rm SM}({1_{G^a_+}, 2_{d_i}, 3_{q_j}, 4_{H^\dagger}})
=-y_\psi g_3(T^a)_{ij}\frac{[41]^2}{[42][43]} 
=y_\psi g_3(T^a)_{ij}\frac{\la 32\ra^2}{\la 12\ra\la 13\ra}\,,
\label{su3case}
\ee
for $SU(3)_c$;
\be
{\cal A}_{\rm SM}({1_{W^a_+}, 2_{e}, 3_{l_j}, 4_{H^\dagger_i}})
=y_e g_2(T^a)_{ij} \frac{[21][41]}{[24][23]} 
=y_e g_2(T^a)_{ij}\frac{\la 23\ra \la 43\ra}{\la 14\ra\la 13\ra}\,,
\label{su2}
\ee
for $SU(2)_L$;
\be
{\cal A}_{\rm SM}({1_{B_+}, 2_{e}, 3_{l}, 4_{H^\dagger}})=y_e g_1\left(
Y_l\frac{[21][41]}{[24][23]} -
Y_{e}\frac{[31][41]}{[34][32]} 
\right)\,.
 \ee
for $U(1)_Y$. 
We also use
$W^a_-B_+|H|^2$  and $W^a_-B_+l\bar{l}$ amplitudes,  that are given by
\be
{\cal A}_{\rm SM}(1_{B_+},2_{H_j},3_{W^a_-},4_{H^\dagger_i})=
g_1g_2Y_H(T^a)_{ij}\, \frac{\la 23\ra \la 43 \ra}{\la 21\ra\la  41\ra}\,.
\label{WBh2}
 \ee
\be
{\cal A}_{\rm SM}(1_{B_+},2_{l_i},3_{W^a_-},4_{\bar{l}_j})=
g_1g_2Y_l(T^a)_{ij}\, \frac{\la 23\ra^2}{\la 21\ra\la  14\ra}\,.
\label{WBl2}
 \ee
All these amplitudes can be determined by just demanding proper transformation
under the little group and  factorization into $n=3$ amplitudes.
 Amplitudes for the opposite helicity, with   particle interchanged with  antiparticle, can
be obtained by complex-conjugating  the above ones.

\section{From the SM EFT Lagrangian  to amplitudes}
\label{app3}

In this Appendix, we provide the relation between our  on-shell  amplitudes
 and   operators used in the common Lagrangian approach  for the SM EFT \cite{Grzadkowski:2010es}.

Let us start with the  dimension-4 operators of the SM EFT. From 
our definition of   the SM gauge couplings, given in Eqs.~(\ref{3pti})--(\ref{3ptint}),
we  find that 
 this  corresponds to take the covariant derivative of a  field  transforming under the fundamental representation of the SM group as
\be
D_\mu=\partial_\mu-i\frac{g_3}{\sqrt{2}}\, T^{a'}G_\mu^{a'}-i\frac{g_2}{\sqrt{2}}\, T^aW_\mu^a-i\frac{g_1}{\sqrt{2}}\, Y_iB_\mu\,,
\label{deco}
\ee 
where the  generators  are normalized as  ${\rm Tr}[T^a T^b]=\delta^{ab}/2$,
and the hypercharge for the Higgs is $Y_H=1/2$. 
Notice
that,  as is usual in amplitude methods  \cite{Dixon:2013uaa}, 
 our gauge couplings carry an extra  $1/\sqrt{2}$, different from  the more common definition of the SM gauge couplings.
One can easily check  that, indeed,  the gauge vertices arising from \eq{deco} lead,  by using \eq{uv} and \eq{eps}, to  Eqs.~(\ref{3pti})--(\ref{3ptint}).

 On the other hand,
our Yukawa coupling defined in \eq{3ptfa} corresponds to that arising  from a  Lagrangian term
\be
-y_e H^\dagger  \bar e_R L_L-y_d H^\dagger  \bar d_R Q_L-y_u \tilde H^\dagger  \bar u_R Q_L=
-y_eH^\dagger el-y_dH^\dagger dq-y_u\tilde H^\dagger uq\,,
\ee
where $\tilde H_i=\epsilon_{ij}H^*_j$, $L_L=(l,0 )^T$ and $\bar e_R=(e,0)$, being $l$ and $e$    Weyl spinors of $h=-1/2$, and similarly for the quarks.

At the dimension-6 level, we have
\begingroup
\allowdisplaybreaks
\bea
\bar L_L\sigma^a\sigma^{\mu\nu}e_R H W^a_{\mu\nu}+h.c.
\ \ \ &\to&\ \ \ \
{\cal A}(1_{\bar e},2_{\bar l_i},3_{W^a_+},4_{H_j})=2\sqrt{2}(\sigma^a)_{ij}[31][32]\,,\\
\ \ \ &\to&\ \ \ \
{\cal A}(1_{ e},2_{l_i},3_{W^a_-},4_{H^\dagger_j})= 2\sqrt{2}(\sigma^a)_{ij}\la 31\ra \la 32\ra\,,\\
\bar L_L\sigma^{\mu\nu}e_R H B_{\mu\nu}+h.c.
\ \ \ &\to&\ \ \ \
{\cal A}(1_{\bar e},2_{\bar l_i},3_{B_+},4_{H_i})=2\sqrt{2}\ [31][32]\,,\\
\ \ \ &\to&\ \ \ \
{\cal A}(1_{ e},2_{l_i},3_{B_-},4_{H^\dagger_i})= 2\sqrt{2}\ \la 31\ra \la 32\ra\,,\\
(\bar L_{L\, i}  e_R) (\bar Q_{L\, j} u_R)\epsilon_{ij}+h.c.
\ \ \ &\to&\ \ \ \
{\cal A}(1_{\bar e},2_{\bar l_i},3_{\bar u},4_{\bar q_j})= [12][34]\epsilon_{ij}\,,\\
\ \ \ &\to&\ \ \ \
{\cal A}(1_{e},2_{ l_i},3_{ u},4_{ q_j})=\la 12\ra \la 34\ra\epsilon_{ij}\,,\\
(\bar L_{L\, i}  u_R) (\bar Q_{L\, j} e_R)\epsilon_{ij}+h.c.
\ \ \ &\to&\ \ \ \
{\cal A}(1_{\bar e},2_{\bar l_i},3_{\bar u},4_{\bar q_j})=- [14][32]\epsilon_{ij}\,,\\
\ \ \ &\to&\ \ \ \
{\cal A}(1_{e},2_{ l_i},3_{u},4_{ q_j})=-\la 14\ra \la 32\ra\epsilon_{ij}\,,\\
W^a_{\mu\nu} W^{a\, \mu\nu} |H|^2 
\ \ \ &\to&\ \ \ \
{\cal A}(1_{W_-^a},2_{W_-^a},3_{H_i},4_{H_i^\dagger})=-2! \la 12\ra^2\,,\\
\ \ \ &\to&\ \ \ \
{\cal A}(1_{W_+^a},2_{W_+^a},3_{H_i},4_{H_i^\dagger})=-2!  [12]^2\,,\\
W^a_{\mu\nu} B^{\mu\nu} H^\dagger \sigma^a H 
\ \ \ &\to&\ \ \ \
{\cal A}(1_{W_-^a},2_{B_-},3_{H_j},4_{H^\dagger_i})=-(\sigma^a)_{ij}\la 12\ra^2\,,\\
\ \ \ &\to&\ \ \ \
{\cal A}(1_{W_+^a},2_{B_+},3_{H_j},4_{H^\dagger_i})=-(\sigma^a)_{ij}  [12]^2\,,\\
W^{a\,\nu}_{\mu}W^{b\, \rho}_{\nu}W^{a\, \mu}_{\rho} f^{abc}
\ \ \ &\to&\ \ \ \
{\cal A}(1_{W_-^a},2_{W_-^b},3_{W_-^c})= i(3!/\sqrt{2}) \la 12\ra\la 23\ra\la 31\ra f^{abc}\,,\\
\ \ \ &\to&\ \ \ \
{\cal A}(1_{W_+^a},2_{W_+^b},3_{W_+^c})=-i(3!/\sqrt{2}) [12][23] [31] f^{abc}
\,.
\eea
\endgroup
The above formulas allow to relate the Wilson coefficients with the coefficients of the on-shell amplitudes that were used in this article.

\section{Dimension-5 operators and their corresponding on-shell amplitudes}
\label{app4}

Similarly as with the amplitudes at order $E^2/\Lambda^2$ (associated to dimension-6 operators), we can determine the extra contributions to amplitudes at order $E/\Lambda$.
These  are given by
 \begin{itemize}
\item 
{\bf n=3, h=-2:} 
  \be
 {\cal A}_{F^2\phi}(1_{V_-},2_{V_-},3_{\phi})=\frac{C_{F^2\phi}}{\Lambda}\la 12\ra^2
 \ , \ \ \ \ \
 {\cal A}_{F\psi^2}(1_{V_-},2_\psi,3_{\psi})=\frac{C_{F\psi^2}}{\Lambda}\la 12\ra \la 13\ra\,.
\label{Fpsi2}
 \ee
\item 
{\bf n=4, h=-1:}
  \be
 {\cal A}_{\psi^2 \phi^2}(1_\psi,2_\psi,3_{\phi},4_\phi)=\frac{C_{\psi^2 \phi^2}}{\Lambda}\la 12\ra\,.
\label{majorana}
\ee
\item 
{\bf n=5, h=0:} 
  \be
 {\cal A}_{\phi^5}(1_\phi,2_\phi,3_\phi,4_\phi,5_\phi)=\frac{C_{\phi^5}}{\Lambda}\,.
\ee
\end{itemize}
In the SM,  only \eq{majorana} is allowed by the gauge symmetry for $\psi=l$,
and it violates  lepton number by two units.
  
One can show that \eq{MF} can also be applied to calculate the anomalous dimensions
of the coefficients   of these amplitudes.
{The proof goes as for the $E^2/\Lambda^2$ case:
we know that \eq{CH} applies to any operator, so we can use it in the limit $Q\to 0$ to get
\eq{MF}.
We only have to be careful with  potential extra contributions present in \eq{CH} 
 that are not considered in \eq{MF}. These are the ones involving $n=3$ amplitudes. 
In particular, they could be relevant in the renormalizations 
 ${\cal A}_{F^2\phi}\to {\cal A}_{\phi^5}$ and ${\cal A}_{F\psi^2}\to {\cal A}_{\psi^2\phi^2}$.
However, one can check that, 
in these one-loop renormalizations, all
triangles and boxes lead to the same integrals as those discussed in Fig.~\ref{deltan1},
and so we can use the conclusions of Appendix \ref{app1} also here, to claim that the absence 
of IR divergencies imposes a cancellation of boxes and triangles
in \eq{MF}. 
Therefore \eq{MF}  must coincide with \eq{CH} in the limit $Q\to 0$.}


\end{document}